\documentclass[12pt]{article}
\usepackage{amsmath,amsthm,latexsym,amssymb,amsfonts,epsfig}

\makeatletter
\@addtoreset{equation}{section}
\makeatother

\newtheorem{Lemma}{Lemma}[section]

\def\be{\begin{equation}}
\def\ee{\end{equation}}
\def\ba{\begin{eqnarray}}
\def\ea{\end{eqnarray}}
\def\Nl{{\mathchoice
{\setbox0=\hbox{$\displaystyle\rm N$}\hbox{\hbox to0pt
{\kern0.4\wd0\vrule height0.9\ht0\hss}\box0}}
{\setbox0=\hbox{$\textstyle\rm N$}\hbox{\hbox to0pt
{\kern0.4\wd0\vrule height0.9\ht0\hss}\box0}}
{\setbox0=\hbox{$\scriptstyle\rm N$}\hbox{\hbox to0pt
{\kern0.4\wd0\vrule height0.9\ht0\hss}\box0}}
{\setbox0=\hbox{$\scriptscriptstyle\rm N$}\hbox{\hbox to0pt
{\kern0.4\wd0\vrule height0.9\ht0\hss}\box0}}}}
\def\Zl{{\mathchoice
{\setbox0=\hbox{$\displaystyle\rm Z$}\hbox{\hbox to0pt
{\kern0.4\wd0\vrule height0.9\ht0\hss}\box0}}
{\setbox0=\hbox{$\textstyle\rm Z$}\hbox{\hbox to0pt
{\kern0.4\wd0\vrule height0.9\ht0\hss}\box0}}
{\setbox0=\hbox{$\scriptstyle\rm Z$}\hbox{\hbox to0pt
{\kern0.4\wd0\vrule height0.9\ht0\hss}\box0}}
{\setbox0=\hbox{$\scriptscriptstyle\rm Z$}\hbox{\hbox to0pt
{\kern0.4\wd0\vrule height0.9\ht0\hss}\box0}}}}
\def\Ql{{\mathchoice
{\setbox0=\hbox{$\displaystyle\rm Q$}\hbox{\hbox to0pt
{\kern0.4\wd0\vrule height0.9\ht0\hss}\box0}}
{\setbox0=\hbox{$\textstyle\rm Q$}\hbox{\hbox to0pt
{\kern0.4\wd0\vrule height0.9\ht0\hss}\box0}}
{\setbox0=\hbox{$\scriptstyle\rm Q$}\hbox{\hbox to0pt
{\kern0.4\wd0\vrule height0.9\ht0\hss}\box0}}
{\setbox0=\hbox{$\scriptscriptstyle\rm Q$}\hbox{\hbox to0pt
{\kern0.4\wd0\vrule height0.9\ht0\hss}\box0}}}}
\def\Rl{{\mathchoice
{\setbox0=\hbox{$\displaystyle\rm R$}\hbox{\hbox to0pt
{\kern0.4\wd0\vrule height0.9\ht0\hss}\box0}}
{\setbox0=\hbox{$\textstyle\rm R$}\hbox{\hbox to0pt
{\kern0.4\wd0\vrule height0.9\ht0\hss}\box0}}
{\setbox0=\hbox{$\scriptstyle\rm R$}\hbox{\hbox to0pt
{\kern0.4\wd0\vrule height0.9\ht0\hss}\box0}}
{\setbox0=\hbox{$\scriptscriptstyle\rm R$}\hbox{\hbox to0pt
{\kern0.4\wd0\vrule height0.9\ht0\hss}\box0}}}}
\def\Cl{{\mathchoice
{\setbox0=\hbox{$\displaystyle\rm C$}\hbox{\hbox to0pt
{\kern0.4\wd0\vrule height0.9\ht0\hss}\box0}}
{\setbox0=\hbox{$\textstyle\rm C$}\hbox{\hbox to0pt
{\kern0.4\wd0\vrule height0.9\ht0\hss}\box0}}
{\setbox0=\hbox{$\scriptstyle\rm C$}\hbox{\hbox to0pt
{\kern0.4\wd0\vrule height0.9\ht0\hss}\box0}}
{\setbox0=\hbox{$\scriptscriptstyle\rm C$}\hbox{\hbox to0pt
{\kern0.4\wd0\vrule height0.9\ht0\hss}\box0}}}}
\def\Hl{{\mathchoice
{\setbox0=\hbox{$\displaystyle\rm H$}\hbox{\hbox to0pt
{\kern0.4\wd0\vrule height0.9\ht0\hss}\box0}}
{\setbox0=\hbox{$\textstyle\rm H$}\hbox{\hbox to0pt
{\kern0.4\wd0\vrule height0.9\ht0\hss}\box0}}
{\setbox0=\hbox{$\scriptstyle\rm H$}\hbox{\hbox to0pt
{\kern0.4\wd0\vrule height0.9\ht0\hss}\box0}}
{\setbox0=\hbox{$\scriptscriptstyle\rm H$}\hbox{\hbox to0pt
{\kern0.4\wd0\vrule height0.9\ht0\hss}\box0}}}}
\def\Ol{{\mathchoice
{\setbox0=\hbox{$\displaystyle\rm O$}\hbox{\hbox to0pt
{\kern0.4\wd0\vrule height0.9\ht0\hss}\box0}}
{\setbox0=\hbox{$\textstyle\rm O$}\hbox{\hbox to0pt
{\kern0.4\wd0\vrule height0.9\ht0\hss}\box0}}
{\setbox0=\hbox{$\scriptstyle\rm O$}\hbox{\hbox to0pt
{\kern0.4\wd0\vrule height0.9\ht0\hss}\box0}}
{\setbox0=\hbox{$\scriptscriptstyle\rm O$}\hbox{\hbox to0pt
{\kern0.4\wd0\vrule height0.9\ht0\hss}\box0}}}}

\title{{\sf Algebraic Quantum Gravity (AQG)\\ III. Semiclassical 
Perturbation Theory}}
\author{\sf 
K. Giesel\thanks{{\sf gieskri@aei.mpg.de, kgiesel@perimeterinstitute.ca}} ~~ and ~
T. 
Thiemann\thanks{{\sf 
thiemann@aei.mpg.de,tthiemann@perimeterinstitute.ca}}\\
\\
{\sf MPI f. Gravitationsphysik, Albert-Einstein-Institut,} \\
           {\sf Am M\"uhlenberg 1, 14476 Potsdam, Germany}\\
\\
{\sf and}\\
\\
{\sf Perimeter Institute for Theoretical Physics,}\\ 
{\sf 31 Caroline Street N, 
Waterloo, ON N2L 2Y5, Canada}}
\date{{\small\sf Preprint AEI-2006-060}}

\begin{document}

\maketitle

\begin{abstract}
{\sf
In the two previous papers of this series we defined a new 
combinatorical 
approach to quantum gravity, Algebraic Quantum Gravity (AQG). We showed 
that AQG reproduces the 
correct infinitesimal dynamics in the semiclassical limit, provided 
one incorrectly 
substitutes the non -- Abelean group $SU(2)$ by the Abelean group 
$U(1)^3$ in the calculations.  

The mere reason why that substitution was performed at all is that in 
the non -- Abelean case the volume operator, pivotal for the definition 
of the dynamics, is not diagonisable by analytical methods. This, in 
contrast to the Abelean case, so far prohibited semiclassical 
computations.

In this paper we show why this unjustified substitution nevertheless 
reproduces the correct physical result:\\
Namely, we introduce for the first time semiclassical perturbation 
theory within AQG (and LQG) which allows to compute expectation values
of interesting operators such as the master constraint as a power series 
in $\hbar$ {\it with error control}. That is, in particular matrix 
elements of fractional powers of the volume operator can be computed     
with extremely high precision for sufficiently large power of 
$\hbar$ in the $\hbar$ expansion.

With this new tool, the non -- Abelean calculation, although technically 
more involved, is then exactly analogous to the Abelean calculation, 
thus justifying the Abelean analysis in retrospect. The results of this 
paper turn AQG into a calculational discipline.}
\end{abstract}

\newpage

\section{Introduction}
\label{s1}

In the two previous companion papers of this series \cite{I,II} we 
introduced a new combinatorial approach to quantum gravity, called 
Algebraic Quantum Gravity (AQG) which uses 
ideas from Loop Quantum Gravity (LQG) \cite{books,reviews}. One of the 
advantages of AQG over LQG is that semiclassical tools for background 
independent quantum field theories already available in the literature 
\cite{GCS,ITP,STW,Complexifier} can be applied also to operators 
encoding the quantum 
dynamics while in LQG this has so far been possible only for kinematical 
operators. The difficulty, as explained in detail in \cite{I} has to do 
with the fact that in LQG the Hamiltonian or Master constraint operator 
\cite{QSD,M1,M2,Test} necessarily changes the number of degrees of 
freedom 
on which the semiclassical state, that it acts on, depends. The 
fluctuations of the degrees 
of freedom added by the operator are therefore not suppressed by the 
semiclassical state and their 
semiclassical behaviour is correspondingly bad. In AQG on the other hand
the dynamics never changes the number of degrees of freedom and the just 
mentioned problem disappears. Furthermore, 
the annoying graph dependence of the normalisable coherent states of LQG
disappears in AQG.

In the companion paper \cite{II} we have displayed a non trivial 
semiclassical calculation which shows that the Euclidean part of the 
Master Constraint for pure gravity in AQG reduces to its classical 
counter part in the $\hbar\to 0$ limit. The same calculation 
demonstrates that its fluctuations and quantum corrections are small.
It is trivially extendable to arbitrary matter coupling and the 
Lorentzian constraint.
However, the calculation was done using an, a priori physically 
unjustified, technical modification of the Master Constraint: namely we 
substituted the correct non -- Abelean gauge group $SU(2)$ of the 
canonical formulation of General Relativity by the incorrect Abelean 
gauge group $U(1)^3$. The reason for why this was done is that the 
$U(1)^3$
analog of the volume operator \cite{Volume}, without which none of the 
pieces  
of the Master Constraint (geometry and matter) can even be defined,
can be diagonalised analytically. 
This is crucial in order that semiclassical calculations can be carried 
out. 

For the non -- Abelean case on the other hand, except in special 
cases, the volume operator cannot be diagonalised analytically\footnote{The
matrix elements of its fourth power are known in closed form
\cite{Volume}, however, the associated matrices on the invariant
subspaces, while finite dimensional, are not diagonisable by 
quadratures.}, which 
prohibited so far any explicit calculations involving the quantum 
dynamics. This is one of the many criticisms that have been spelled out 
recently in \cite{Nicolai}. 

In the present paper we show that this criticism, as many others, is 
unjustified, thereby introducing 
{\it semiclassical perturbation theory} for AQG and LQG. Basically, what 
we are interested in are expectation values or matrix elements of 
(powers of) the Master Constraint Operator in semiclassical states.
We will show how to compute those as a power series in $\hbar$ 
{\it with error control!} Notice that in usual perturbative QFT 
there is no error control. In fact, there the perturbation series is 
known to be only asymptotic for all realistic theories.   

As we will see, and as should be expected 
from \cite{I,II}, the non -- trivial part of the 
corresponding $\hbar$ expansion consists in the evaluation of the 
matrix elements of fractional powers of the volume operator. The basic 
mathematical physics tools that we employ here are the spectral theorem 
for self -- adjoint 
operators and the directed set structure of the cone of positive 
operators. More is not needed in order to define the series. The fact 
that makes the errors small on the other hand relies on the 
phantastically good semiclassical properties of the semiclassical states 
developed in \cite{GCS}. For instance, if we are interested in 
cosmological questions, the power series is typically in terms of the
classicality parameter $t=\Lambda \hbar G\approx 10^{-120}$ where 
$\Lambda$ is the cosmological constant and $G$ is Newton's constant.
This should be contrasted with the Feinstrukturkonstante $\alpha\approx 
1/137$ of QED. Here the extreme weakness of the gravitational 
interaction comes to help. Notice that our power expansion, for 
cosmological settings, is as fast converging as for the new spin 
foam model introduced in \cite{FS}.\\
\\
This article is organised as follows:\\

In section two we develop systematically semiclassical perturbation 
theory for AQG. For the benefit of the reader not interested in the 
detailed proofs, let us just state that if we are interested in the matrix 
elements $<\psi,V_v^{2q} \psi'>$ of the volume operator $V_v$ associated 
with a given vertex $v$ which is not calculable analytically, then we can 
replace $V_v^{2q}$ within the matrix element, up to $\hbar^{k+1}$ 
corrections, by the calculable quantity 
\be \label{1.1}
<\psi,Q_v\psi>^{2q}[1+\sum_{n=1}^{2k+1} \;(-1)^{n+1}\;
\frac{q(1-q)..(2k-q)}{n!}\;\; (\frac{Q_v^2}{<\psi,Q_v\psi>^2}-1)^n]
\ee
where $Q_v$ is related to the volume by $V_v=\root 4\of{Q_v^2}$. The 
operator $Q_v$ is a polynomial in flux operators and its matrix elements
are known in closed form.

In section three we show how the results of \cite{GCS} can be used 
in order to perform concrete calculations (numbers!) using the tools of 
section two. This will show semiclassical perturbation theory {\it at 
work}!

In section four we conclude and explain why the Abelean results of 
\cite{II} carry over qualitatively and quantitatively to the non -- 
Abelean case as far as the zeroth order in $\hbar$ are concerned while 
there are finite differences in the first and higher order corrections. 
This justifies the calculation of \cite{II} and demonstrates that AQG is a 
calculational discipline.

\section{Semiclassical Perturbation Theory}
\label{s2}

\subsection{The Idea}
\label{s2.1}

In semiclassical applications of AQG we are interested, in particular, 
in expectation 
values of (finite powers of) the Master Constraint with respect to 
coherent states $\psi$. As displayed explicitly in \cite{I} the 
operators of which we have to take expectation values are then 
linear combinations of expressions of the form (symbolically)
\be \label{2.1}
<\psi, p_1(h) F_1(V_{v_1}) p_2(h) F_2(V_{v_2}) ..  F_N(V_{v_N}) 
p_{N+1}(h)\psi>
\ee
where $p_j$ are certain polynomials in the holonomies along edges or 
loops adjacent to the vertices $v_1,..,v_N$ and 
$F_I$ are certain functions of the volume operator $V_{v_I}$ of the form
$F_I(V_{v_I})=(Q_{v_I}^2)^{q_I}$. Here $0<q_I=m_I/n_I \le 1/4$ is 
a rational number and $m_I,n_I$ are relative prime. The self -- 
adjoint operator $Q_v$ is explicitly given by\footnote{For convenience we 
are working here 
with the dimensionless volume operator that results from the actual one 
by dividing it by $a^3$, where $a$ is the length scale parameter that 
enters the definition of the coherent states.} 
\be \label{2.2}
Q_v=i t^3\epsilon_{jkl} 
\; \frac{X^j_{e_{1+}(v)}-X^j_{e_{1-}(v)}}{2}   
\; \frac{X^k_{e_{2+}(v)}-X^k_{e_{2-}(v)}}{2}   
\; \frac{X^l_{e_{3+}(v)}-X^l_{e_{3-}(v)}}{2}   
\ee
\\
where we have used the notation from \cite{I,II}: By $e_{I\sigma}(v)$ we 
mean the edge of a cubic algebraic graph outgoing from vertex $v$ into 
the positive ($\sigma=+$) or negative ($\sigma=-$) direction 
respectively and $X^j_I$ denotes the corresponding right invariant 
vector field of $SU(2)$. Notice that $e_{I+}(v)=e_I(v),\;
e_{I-}(v)=[e_I(v-\hat{I})]^{-1}$. The classicality parameter is 
$t=\ell_P^2/a^2$ where $a$ is some length scale enters the definition of 
the coherent states.\\ 
\\
The naive idea to compute expectation values of (\ref{2.2}) is as 
follows: Consider 
\be \label{2.3}
x_I:=\frac{Q_{v_I}^2}{<\psi,Q_{v_I}\psi>^2}-1
\ee
The operator $x_I$ is bounded from below, $x_I\ge -1$ and the quantity 
$<\psi,Q_{v_I}\psi>$ can be computed exactly by the methods of 
\cite{GCS}. We have 
\be \label{2.4}
F_I(V_{v_I})=|<\psi,Q_{v_I}\psi>|^{2q_I} \;f_I(x_I),\;\;
f_I(x_I)=(1+x_I)^{q^I}
\ee 
The idea is now to use the power expansion of the function 
$t\mapsto f(t)=(1+t)^q,\;-1\le t<\infty$ given by 
\be \label{2.5}
f(t):=1+\sum_{n=1}^\infty 
\left( \begin{array}{c} q \\ n \end{array} \right) t^n,\;
\left( \begin{array}{c} q \\ n \end{array} \right)=(-1)^{n+1}
\frac{q(1-q)..(n-1+q)}{n!}
\ee
and to use the spectral theorem in order to get an expansion of 
$F_I(V_{v_I})$ in terms of $x_I$ of which coherent state matrix elements 
are computable by the methods of \cite{GCS}, specifically
\ba \label{2.6}
f_I(x_I)
&=& \int_{-1}^\infty f_I(t) dE_I(t)
=\int_{-1}^\infty [1+\sum_{n=1}^\infty 
\left( \begin{array}{c} q \\ n \end{array} \right) t^n]\;
dE_I(t)
\nonumber\\
&=&  [1+\sum_{n=1}^\infty 
\left( \begin{array}{c} q \\ n \end{array} \right) x_I^n]\;
\ea
where $E_I$ is the projection valued measure associated with $x_I$. 
However, the second equality is wrong because the power expansion 
(\ref{2.6}) does not converge outside the open interval $t\in (-1,1)$.
Hence, this naive idea is false and must be substituted by a 
rigorous argument. \\
\\
Much of the effort that follows will be devoted to showing that the 
power expansion (\ref{2.5}) can nevertheless be used in order to get 
reliable numerical results for the intended expectation value 
calculations including error estimates. More precisely, the power 
expansion up to order $2k+1$ gives reliable values for the matrix 
elements including $\hbar^k$ corrections while the remainder can be 
estimated to be finite and of order $\hbar^{k+1}$. The proof of this 
fact is 
somewhat involved and we therefore split it into several subsections.
The reader not interested in the proof can jump directly to section 
\ref{s3}. 

\subsection{Some Basic Tools} 
\label{s2.2}

In this subsection we will state and prove some basic lemmas which will 
be employed over and over again in the core part of the proof.
\begin{Lemma} \label{la2.1} ~\\
For each $k\ge 0$ there exists $0<\beta_k<\infty$ such that 
\be \label{2.7}
f_{2k+1}(t)-\beta_k t^{2k+2} \le f(t) \le f_{2k+1}(t)
\ee
where $f_k(t)$ denotes the partial Taylor series of $f(t)=(1+t)^q,\;
0<q\le 1/4$ up to order $t^k$.
\end{Lemma}
\begin{proof}
By Taylor's theorem we have 
\be \label{2.8}
f(t)=f_k(t)+R_k(t),\; f_k(t)=1+\sum_{n=1}^k 
\left( \begin{array}{c} q \\ n \end{array} \right) t^n,\;
R_k(t)=\int_0^t ds \frac{f^{(k+1)}(s)}{k!} (t-s)^k
\ee
for any $-1<t<\infty$. 

Since 
$f^{(2k+2)}(s)=-q(1-q)..(2k+1-q)(1+s)^{q-2k-2}\le 0$ for all $s>-1$
and $(t-s)^{2k+1}$ is positive (negative) for $t\ge s \ge 0$ ($-1\le 
t\le s \le 0$) it follows that $R_{2k+1}(t)\le 0$ for all $t>-1$, hence
$f_{2k+1}(t)\ge f(t)$ for all $t> -1$ and also for $t=-1$ by continuity.   

On the other hand, $f^{(2k+3)}(s)=q(1-q)..(2k+2-q)(1+s)^{q-2k-3}\ge 0$ 
for all $s>-1$. Since $(t-s)^{2k+2}\ge 0$ for all $s$ we certainly have 
$f_{2k+2}(t)\le f(t)$ for $t\ge 0$ and
$f_{2k+2}(t)\ge f(t)$ for $-1\le t\le 0$. Now 
$f_{2k+2}(t)=f_{k+1}(t)+
\left( \begin{array}{c} q \\ 2k+2 \end{array} \right) t^{k+2}$. Then
for any $\beta_k> -
\left( \begin{array}{c} q \\ 2k+2 \end{array} \right)$
we still have $f_{2k+1}(t)-\beta_k t^{2k+2} \le f(t)$ for $t\ge 0$.
Moreover, since $0\le f_{2k+1}(t)-f(t)$ is bounded from above for 
$-1 <t\le 0$ there exists a finite $\beta_k$ such that also 
$f_{2k+1}(t)-\beta_k t^{2k+2} \le f(t)$ for $-1\le t\le 0$.
\end{proof}
The optimal (lowest) value of $\beta_k$ is not easy to obtain by 
analytical methods but certainly by numerical methods. As an example, let 
us determine it 
analytically for the case $k=0$.
\begin{Lemma} \label{la2.2} ~\\
\be \label{b.7}
1+qt -(1-q) t^2\le (1+t)^q \le 1+q t
\ee
for all $t\ge -1$ and all $0<q<1$. 
\end{Lemma}
\begin{proof}
Consider $f(t)=1+qt-(1+t)^q$. We have 
$f'(t)=q(1-(1+t)^{q-1})$ so that $f'\ge 0$ for $t\ge 0$ and $f'\le 0$ 
for
$-1\le t\le 0$. Thus, $f(0)=0$ is the absolute minimum of $f$.
Let now $g(t)=f(t)-\beta t^2$. In order that $g\le 0$ we certainly need 
$g(-1)=1-q-\beta\le 0$ so $\beta\ge 1-q$. We have
\be \label{b.8}
g'=f'-2\beta t=q[\frac{1-\frac{1}{y^{1-q}}}{y-1}-p](y-1)
\ee
where $y=1+t\ge 0$ and $p=2\beta/q\ge 2(1-q)/q>1-q$. Consider 
\be \label{b.9} 
h(y)=\frac{1-\frac{1}{y^{1-q}}}{y-1}\;\; \Rightarrow
h'(y)=\frac{1}{y^{2-q} (y-1)^2}k(y),\;\;k(y)=(1-q)(y-1)-y^{2-q}+y
\ee
It follows that $k'=(2-q)(1-y^{1-q})$ hence $y=1$ is the maximum of $k$ at 
which $k(1)=0$. Hence $k\le 0$ and thus $h'\le 0$ is strictly monotonously 
decreasing. Since $h(1)=1-q$ and $p>1-q$ it follows that $g'$ is strictly 
monotonously decreasing for $y>1$. The equation $g'=0$ or $h=p$ has one 
more solution $0<y_{q,p}<1$ because $h(0)=\infty$ and $h$ is decreasing.
It follows that $g'\le 0$ for $0\le y\le y_{q,p}$ and $y\ge 1$ and 
$g'\ge 0$ for $y_{q,p}\le y\le 1$. Hence $y_{q,p}$ is a local minimum of 
$g$ and $y=0,1$ are local maxima at which 
$g(x=-1)=1-q-\beta,\;g(x=0)=0$.
Thus, in order that $g\le 0$ it is necessary and sufficient that 
$\beta\ge 
1-q$ where $\beta=1-q$ is the sharpest bound.
\end{proof}
To see what lemma \ref{la2.1} is good for we state the following 
simple lemma.
\begin{Lemma} \label{la2.3} ~\\
Let $B_- \le B \le B_+$ be self -- adjoint operators and set 
$\bar{B}:=[B_+ + B_-]/2,\;\;\Delta B:=[B_+ - B_-]/4$. Then for any 
states $\psi_1,\;\psi_2$ in the common domain of all three operators we 
have
\be \label{b.0}
|\Re(<\psi_1,[B-\bar{B}]\psi_2>)|,\;
|\Im(<\psi_1,[B-\bar{B}]\psi_2>)| \le \;\;
<\psi_1,[\Delta B]\psi_1>+<\psi_1,[\Delta B]\psi_1>
\ee
\end{Lemma}
\begin{proof}
Consider the the polarisation identity for any self adjoint operator $B$
\be 
\label{b.1}
<\psi_1,B\psi_2>=\frac{1}{4}\sum_{\iota^4=1} \bar{\iota} 
<\psi_\iota, 
B\psi_\iota>
\ee
where $\psi_\iota=\psi_1+\iota \psi_2$. By assumption the operators
$B_+ - B,\;B- B_-$ are positive. Thus 
\ba \label{b.2}
&&\frac{<B_->_{+1}-<B_+>_{-1}}{4}   
\le
\Re(<\psi_1,B\psi_2>)=
\frac{<B>_1-<B>_{-1}}{4}\le   
\frac{<B_+>_1-<B_->_{-1}}{4}   
\nonumber\\
&&\frac{<B_->_{-i}-<B_+>_{+i}}{4}   
\le
\Im(<\psi_1,B\psi_2>)=
\frac{<B>_{-i}-<B>_{+i}}{4}\le   
\frac{<B_+>_{-i}-<B_->_{+i}}{4}   
\nonumber\\
&& 
\ea
where $<B>_\iota:=<\psi_\iota,B \psi_\iota>$ etc. 
We conclude 
\ba \label{b.3}
|\Re(<\psi_1,[B-\bar{B}]\psi_2>)| &\le & 
<\psi_1,[\Delta B] \psi_1> +<\psi_2,[\Delta B]\psi_2>
\nonumber\\
|\Im(<\psi_1,[B-\bar{B}]\psi_2>)| &\le & 
<\psi_1,[\Delta B] \psi_1> +<\psi_2,[\Delta B]\psi_2>
\ea
as claimed.
\end{proof}
The idea behind this lemma is that, even if we cannot compute the 
matrix elements of $B$, provided we 
can compute the matrix elements of $\bar{B}$ and $\Delta{B}$, formula
(\ref{b.3}) shows that the matrix element of $\bar{B}$ is a good 
approximation to that of $B$ provided that the respective expectation 
values of $\Delta B$ are small.  

As an example, consider $B=(Q^2)^q$ with $Q$ as in (\ref{2.2}) 
polynomial in the elementary operators (flux operators for GR) so that
matrix elements can be computed by the methods of \cite{GCS}. 
We have for 
any $\lambda>0$ that $B=\lambda^q[1+(Q^2-\lambda)/\lambda]^q$ and 
obviously
$x=(X-\lambda)/\lambda\ge -1$. Then we may apply the above lemma 
and set $B_+=\lambda^q[1+x],\;B_-=\lambda^q[1+qx-(1-q)x^2]$. Given a 
coherent state $\psi_1$ we could set $\lambda=<\psi_1,Q^2\psi_1>$ so 
that $<\psi_1,B_+\psi_1>=\lambda^q$ and  
$<\psi_1,B_-\psi_1>=\lambda^q
[1-(1-q)(\frac{<\psi_1,Q^4\psi_1>}{\lambda^2}-1)]$.
Hence, the expectation value of $B=(Q^2)^q$ is approximately given by 
$<Q^2>^q$ up to a correction which is controlled by the fluctuation of 
$Q^2$ as one would expect. Notice that it has been established \cite{GCS} that 
$<Q^2>$ agrees with the classical value to lowest order in $\hbar$.

\subsection{The Expansion}
\label{s2.3}

Turning to the general matrix element of interest (\ref{2.1}) we expand 
\ba \label{b.9a}
&& <\psi, p_1(h) F_1(V_{v_1}) p_2(h) F_2(V_{v_2}) ..  F_N(V_{v_N}) 
p_{N+1}(h)\psi>
\nonumber\\
&=& \prod_{j=1}^n F_j^0 
<\psi, p_1 [1+f_1] p_2 [1+f_2] ..  [1+f_N] p_{N+1}\psi>
\nonumber\\
&=& \prod_{j=1}^n F_j^0 
[<\psi, p_1 p_2.. p_{N+1}\psi>+R]
\ea
where $F_j^0=<\psi, Q_{v_j} \psi>^{2q_j}$ and  
the remainder $R$ is a linear combination of terms of the 
form 
\be \label{b.10}
<\psi,p'_1 f'_1 .. f'_l p'_{l+1} \psi>
\ee
with $l=1,..,N$ and $p'_j$ are products of the $p_j$ while each $f'_j$ 
coincides with one of the $f_k$. \\
\\
Hence we are left with computing (\ref{b.10}). 
To do this, we use lemma \ref{la2.1} and find optimal $\beta_{j,k}$ such that 
\be \label{b.11}
f_j^-(t):=f_{j,2k+1}(t)-\beta_{j,k} t^{2(k+2)} \le f_j(t) \le 
f_{j,2k+1}(t) =: f^+_j(t)
\ee
for all $t\ge -1$. We will now iterate lemma \ref{la2.3} frequently. In 
what follows we just consider the real part of (\ref{b.10}), the 
imaginary part works analogously as displayed in (\ref{b.0}).\\
\\
We start with 
\ba \label{b.12}  
&& <\psi,p_1 f_1 .. f_N p_{N+1} \psi>
\nonumber\\
&=& 
<[\overline{p_{[(N+1)/2]}} f_{[(N+1)/2]-1} .. f_1 \overline{p_1} \psi],
f_{[(N+1)/2]} [p_{[(N+1)/2]-1} f_{[(N+1)/2]+1} .. f_N p_{N+1} \psi]>
\nonumber\\
&=:& <\psi_1,f_{[(N+1)/2]} \psi_2>
\ea
Here $[t]$ denotes the Gauss bracket of the real number $t$.
Application of (\ref{b.3}) reveals that $|R-R_1|\le R_2 +R_3$ where  
\ba \label{b.13}
R &=& \Re(<\psi,p_1 f_1 .. f_N p_{N+1} \psi>)
\nonumber\\
R_1 &=& \Re(<\psi_1, \bar{f}_{[(N+1)/2]} \psi_2>)
\nonumber\\
R_2 &=& <\psi_1, \Delta{f}_{[(N+1)/2]} \psi_1>
\nonumber\\
R_3 &=& <\psi_2, \Delta{f}_{[(N+1)/2]} \psi_2>
\ea
where $\bar{f}_j=(f_j^+ +f_j^-)/2$
and $\Delta f_j=(f_j^+ -f_j^-)/4$ are positive operators of which 
matrix elements can indeed be computed using the techniques of 
\cite{GCS}. 

Notice that in $R_1$ we have achieved a reduction from $N$ operator
insertions $f_1,..,f_N$ of which matrix elements cannot be computed 
to $N-1$ insertions $f_1,..,f_{[(N+1)/2]-1},f_{[(N+1)/2]+1},..,f_N$. 
However, in $R_1$ we still have $N-2$ ($N-1$) insertions if $N$ is even 
(odd) while in $R_2$ we still have $N$ ($N-1$) insertions if $N$ is odd.
It follows that by iterating (\ref{b.13}) we can never achieve that 
all occurring expressions only contain operator insertions of which 
matrix elements are computable analytically. Notice that 
judiciously we started the iteration with the ``middle'' 
operator, otherwise we would create even more than $N$ operator 
insertions of which matrix elements cannot be computed. By 
choosing the middle, the number of those insertions at least 
does not increase.

While the above procedure therefore never stops at a stage at which 
everything is exactly computable, we will show that after finitely many 
steps one reaches a stage at which the non computable terms can be 
estimated from above by computable terms and those estimates are of 
higher order in $\hbar$ than the order of the power expansion that we 
wanted to achieve. More precisely, we 
will show that 
after finitely many steps we obtain one expression which only involves 
the $\bar{f}_j$ and thus the associated expectation value can be 
computed by the methods of \cite{GCS} and contains $\hbar$ corrections 
up to order $\hbar^k$. In addition there 
are a finite number of terms each of which still contain 
at most $N$ operator insertions of which matrix elements cannot be 
computed plus at least $2l+1$ insertions of operators of the form 
$\Delta f'_j$. These terms appear in the manifestly positive form, say for 
$N$ even
\ba \label{b.14}
&& <P'_1 [\Delta F'_1] P'_2 .. P'_l [\Delta F'_l] P_1 F_1 P_2 .. F_{N/2} 
P_{N/2+1} \psi|
\nonumber\\
&&  [\Delta F'_{l+1}]\;\; |\;\;
P'_1 [\Delta F'_1] P'_2 .. P'_l [\Delta F'_l] P_1 F_1 P_2 .. F_{N/2} 
P_{N/2+1} \psi>
\ea
Here the $F'_j,\; F_k$ are  
elements of the original set $f_1,..,f_N$. Likewise the $P'_j,P_k$ are 
elements of the original set 
$p_1,..,p_{N+1}$.  Instead of $N/2$ insertions 
of type $F_k$ we might also have some insertions of type $\bar{F}_k$ but
since the latter have computable matrix elements, their estimate 
is of an even higher order in $\hbar$ than 
(\ref{b.14}) as we 
will see so that
(\ref{b.14}) provides the type of term whose estimate gives the 
lowest power of $\hbar$. Hence we will show that for $l$ sufficiently 
large
(\ref{b.14}) can be estimated by a computable expression of order at 
least $\hbar^{k+1}$. 

\subsection{Resolutions of Unity}
\label{s2.4}

To estimate (\ref{b.14}) we use the 
overcompleteness property of our coherent states \cite{GCS} and insert 
resolutions 
of unity to cast (\ref{b.14}) into the form 
\ba \label{b.15}
&& 
\int d\nu_{1,1} .. \int d\nu_{1,N/2} 
\int d\nu_{2,1} .. \int d\nu_{2,N/2+1} 
\int d\nu'_{1,1} .. \int d\nu'_{1,l} 
\int d\nu'_{2,1} .. \int d\nu'_{2,l}
\times \nonumber\\
&& \times
\int d\nu_{3,1} .. \int d\nu_{3,N/2} 
\int d\nu_{4,1} .. \int d\nu_{4,N/2+1} 
\int d\nu'_{3,1} .. \int d\nu'_{3,l} 
\int d\nu'_{4,1} .. \int d\nu'_{4,l}
\times \nonumber\\
&& \times  
\overline{
\prod_{j=1}^{l-1}   
<\psi'_{1,j}, [\Delta F'_j] \psi'_{2,j+1}> 
<\psi'_{2,j}, P'_j \psi'_{1,j}>
\;\;\;
<\psi'_{2,l}, P'_l \psi'_{1,l}>
<\psi'_{1,l}, [\Delta F'_l] \psi_{2,1}>}
\times\nonumber\\ && \times 
\overline{
\prod_{k=1}^{N/2} <\psi_{1,k},F_k \psi_{2,k+1}> \; 
<\psi_{2,k}, P_k \psi_{1,k}>
\;\;\;
<\psi_{2,N/2+1},  P_{N/2+1} \psi>
}
\times\nonumber\\
&& \times <\psi'_{2,1}, [\Delta F'_{l+1}] \psi'_{4,1}>
\times\nonumber\\
&&
\times 
\prod_{j=1}^{l-1}   
<\psi'_{3,j}, [\Delta F'_j] \psi'_{4,j+1}> 
<\psi'_{4,j}, P'_j \psi'_{3,j}>
\;\;\;
<\psi'_{4,l}, P'_l \psi'_{3,l}>
<\psi'_{3,l}, [\Delta F'_l] \psi_{4,1}> 
\times\nonumber\\
&& \times 
\prod_{k=1}^{N/2} <\psi_{3,k},F_k \psi_{4,k+1}> \; 
<\psi_{4,k}, P_k \psi_{3,k}>
\;\;\;
<\psi_{4,N/2+1},  P_{N/2+1} \psi>
\ea
Here the measures $\nu_{i,k},\;\nu'_{i,j}$ are basically Liouville 
measures \cite{GCS} over as many copies of the cotangent bundle 
$T^\ast(SU(2))$ as 
there are edges involved in the expectation value under consideration 
and each of the integrals are over the entire corresponding phase 
space.
The states $\psi_{i,k},\;\psi'_{i,j}$ are labelled by the points in 
that phase space, see \cite{GCS} for details.

\subsection{Sketch of the Estimate}
\label{s2.5}

Before we give the rigorous argument, let us give its heuristic form
in order to understand what we are driving at:
In order to estimate (\ref{b.15}) we make use of the following fact
that has been proved in \cite{GCS}: 
\be \label{b.16}
<\psi_1,P\psi_2>=
<\psi_1,\psi_2> [E_0(\psi_1,\psi_2)+\hbar E_1(\psi_1,\psi_2)]
\ee
where $\psi_1,\psi_2$ are arbitrary coherent states whose overlap 
function $<\psi_1,\psi_2>$ is sharply peaked (Gaussian) at the point in 
phase 
space by which they are labelled. The functions $E_0,E_1$ are 
absolutely integrable against that Gaussian over both resolution 
measures $\nu_1,\nu_2$ and are of zeroth order in $\hbar$.

Since (\ref{b.16}) is sharply peaked at $\psi_1=\psi_2$ we can basically 
set $E_0(\psi_1,\psi_2)=E_0(\psi_1,\psi_1)=<\psi_1,P\psi_1>$ in the 
essential support of the overlap function, the corrections being of 
higher order in $\hbar$, just like the $E_1$ contribution. This is what 
results after integrating over the measure corresponding to $\psi_2$.
Thus by integrating with respect to the measures 
$\nu_{2,k},\nu'_{2,j},\nu_{4,k},\nu'_{4,j}$ we can simplify (\ref{b.15})
up to $\hbar$ corrections to
\ba \label{b.17}
&& 
\int d\nu_{1,1} .. \int d\nu_{1,N/2} 
\int d\nu'_{1,1} .. \int d\nu'_{1,l} 
\times \nonumber\\
&& \times
\int d\nu_{3,1} .. \int d\nu_{3,N/2} 
\int d\nu'_{3,1} .. \int d\nu'_{3,l} 
\times \nonumber\\
&& \times  
\overline{
\prod_{j=1}^{l-1}   
<\psi'_{1,j}, [\Delta F'_j] \psi'_{1,j+1}> 
<\psi'_{1,j}, P'_j \psi'_{1,j}>
\;\;\;
<\psi'_{1,l}, P'_l \psi'_{1,l}>
<\psi'_{1,l}, [\Delta F'_l] \psi_{1,1}> }
\times\nonumber\\ &&\times \overline{
\prod_{k=1}^{N/2-1} <\psi_{1,k},F_k \psi_{1,k+1}> \; 
<\psi_{1,k}, P_k \psi_{1,k}>
}
\times \nonumber\\ && \times \overline{
<\psi_{1,N/2},F_{N/2} \psi> \; 
<\psi_{1,N/2}, P_{N/2} \psi_{1,N/2}>\;
<\psi,  P_{N/2+1} \psi>
}
\times\nonumber\\
&& \times <\psi'_{1,1}, [\Delta F'_{l+1}] \psi'_{3,1}>
\times\nonumber\\
&&
\times 
\prod_{j=1}^{l-1}   
<\psi'_{3,j}, [\Delta F'_j] \psi'_{3,j+1}> 
<\psi'_{3,j}, P'_j \psi'_{3,j}>
\;\;\;
<\psi'_{3,l}, P'_l \psi'_{3,l}>
<\psi'_{3,l}, [\Delta F'_l] \psi_{3,1}> 
\times\nonumber\\ && \times
\prod_{k=1}^{N/2-1} <\psi_{3,k},F_k \psi_{3,k+1}> \; 
<\psi_{3,k}, P_k \psi_{3,k}>
\times\nonumber\\ && \times
<\psi_{3,N/2},F_{N/2} \psi> \; 
<\psi_{3,N/2}, P_{N/2} \psi_{3,N/2}>\;
<\psi,  P_{N/2+1} \psi>
\nonumber
\ea
In order to estimate (\ref{b.17}) further we will prove that for 
arbitrary coherent states $\psi_1,\psi_2$ we have 
\be \label{b.18}
<\psi_1,[\Delta F'] \psi_2>=\hbar^{k+1} <\psi_1,\psi_2> 
[G'_0(\psi_1,\psi_2)+\hbar 
G'_1(\psi_1,\psi_2)]
\ee
where $G'_0,G'_1$ are absolutely 
integrable with respect to both 
$\nu_1,\nu_2$ and of zeroth order in $\hbar$.
Similarly we will show that
\be \label{b.19}
|<\psi_1,F \psi_2>|\le \hbar^{-3}  
<\psi_1,\psi_2>'\;\; [G_0(\psi_1,\psi_2)+\hbar G_1(\psi_1,\psi_2)]
\ee
where $G_0,G_1$ are both of zeroth order in $\hbar$
and both terms are separately absolutely integrable with respect to both 
$\nu_1,\nu_2$. Notice that while (\ref{b.16}) and (\ref{b.18}) are 
equalities, (\ref{b.19}) is an inequality. It is due to this inequality,
{\it which provides the key part of the proof},
that a negative power of $\hbar$ appears in 
(\ref{b.19}) which is the  
price to pay for the estimate. The estimate is necessary to 
carry out because 
the left hand side is not calculable analytically. We conjecture that
the matrix element (\ref{b.19}) admits a sharper bound which does not 
involve the $\hbar^{-3}$ power, however, the estimate presented below is 
not able to deliver such a result. 

Notice also the prime at the overlap
function $<\psi_1,\psi_2>$ in (\ref{b.19}). This prime is in order to 
indicate that 
$<\psi_1,\psi_2>$ is a Gaussian only with respect to the momentum 
variables of the phase space but not with respect to the position 
variables of six copies of $SU(2)$ corresponding to the six edges 
adjacent to a vertex in the definition of the volume operator. This 
is unproblematic as far as integrability is concerned 
because the configuration space is a compact group and the Haar
measure involved in $\nu$ is normalised. However, the missing damping 
factor leads to an additional negative $\hbar$ power that comes from 
the measures $\nu$ as we will see 
momentarily.  

Notice that the polynomials $P$ are bounded operators
so that $|<\psi, P \psi>|\le ||P||$ where $||P||$ is the sup norm 
on the Abelean $C^\ast-$algebra of functions of connections. Let us 
assume for simplicity that $||P||\le 1$ which is typically the 
case because in our application $P$ is a matrix element of a holonomy 
in some representation of $SU(2)$. Hence we 
can estimate (\ref{b.17}) by
\ba \label{b.20}
&& 
\int d\nu_{1,1} .. \int d\nu_{1,N/2} 
\int d\nu'_{1,1} .. \int d\nu'_{1,l} 
\times \nonumber\\
&& \times
\int d\nu_{3,1} .. \int d\nu_{3,N/2} 
\int d\nu'_{3,1} .. \int d\nu'_{3,l} 
\times \nonumber\\
&& \times  
\prod_{j=1}^{l-1}   
|<\psi'_{1,j}, [\Delta F'_j] \psi'_{1,j+1}>| 
\;\;\;
|<\psi'_{1,l}, [\Delta F'_l] \psi_{1,1}>|
\times\nonumber\\ && \times
\prod_{k=1}^{N/2-1} |<\psi_{1,k},F_k \psi_{1,k+1}>|  
\;\;\;
|<\psi_{1,N/2},F_{N/2} \psi>| 
\times\nonumber\\
&& \times |<\psi'_{1,1}, [\Delta F'_{l+1}] \psi'_{3,1}>|
\times\nonumber\\
&&
\times 
\prod_{j=1}^{l-1}   
|<\psi'_{3,j}, [\Delta F'_j] \psi'_{3,j+1}>| 
\;\;\;
|<\psi'_{3,l}, [\Delta F'_l] \psi_{3,1}>| 
\times\nonumber\\ && \times
|\prod_{k=1}^{N/2-1} <\psi_{3,k},F_k \psi_{3,k+1}>|  
\;\;\;
|<\psi_{3,N/2},F_{N/2} \psi>|  
\nonumber\\
&\le & \hbar^{(2l+1)(k+1)-3N}
\times \nonumber\\
&& \times
\int d\nu_{1,1} .. \int d\nu_{1,N/2} 
\int d\nu'_{1,1} .. \int d\nu'_{1,l} 
\times \nonumber\\
&& \times
\int d\nu_{3,1} .. \int d\nu_{3,N/2} 
\int d\nu'_{3,1} .. \int d\nu'_{3,l} 
\times \nonumber\\
&& \times  
\prod_{j=1}^{l-1}   
|<\psi'_{1,j},\psi'_{1,j+1}> G'_{0j}(\psi'_{1,j+1})| 
\;\;\;
|<\psi'_{1,l},\psi_{1,1}> G'_{0l}(\psi_{1,1})| 
\times\nonumber\\ && \times
\prod_{k=1}^{N/2-1} |<\psi_{1,k},\psi_{1,k+1}>' G_{0k}(\psi_{1,k+1})|  
\;\;\;
|<\psi_{1,N/2},\psi>' G_{0N/2}(\psi)| 
\times\nonumber\\
&& \times |<\psi'_{1,1},\psi'_{3,1}> 
G'_{0l+1}(\psi'_{3,1})|
\times\nonumber\\
&&
\times 
\prod_{j=1}^{l-1}   
|<\psi'_{3,j}, \psi'_{3,j+1}> G'_{0j}(\psi'_{3,j+1})| 
\;\;\;
|<\psi'_{3,l}, \psi_{3,1}> G'_{0l}(\psi_{3,1})| 
\times\nonumber\\ &&\times
|\prod_{k=1}^{N/2-1} <\psi_{3,k},\psi_{3,k+1}>' G_{0k}(\psi_{3,k+1})|  
\;\;\;
|<\psi_{3,N/2},\psi>' G_{0N/2}(\psi)|  
\ea
where $G_{0k}(\psi)=G_{0k}(\psi,\psi)$ etc.
Basically, the remaining integral has as many Gaussians in the
momentum variables as there are measures. However, there are $N$ factors 
of the form $<\psi_1,\psi_2>'$, hence there are $6N$ Gaussians in the 
configuration variables missing. Since the associated $6N$ resolution 
measures come with a factor of $\hbar^{-3}$ for each of which only 
$\hbar^{-3/2}$ gets absorbed into the integration measure of the 
momentum Gaussian integral, the final integral in (\ref{b.20}) is of 
order $\hbar^{-3/2(6N)}=\hbar^{-9N}$. Hence the overall estimate 
is of order $\hbar^{(2l+1)(k+1)-12N}$ at most times a phase space 
point (defined by the state $\psi$) depending factor
whose value is of the same order of 
magnitude as the leading order of $\hbar$ coefficient of the calculable 
quantity
\be \label{b.21}
<\psi,F^+_1..F^+_{N/2} [\Delta F'_1] .. [\Delta F'_{l+1}] \psi>
\ee
From this heuristic reasoning we 
conclude that in order that the correction terms be of order 
$\hbar^{k+1}$ we must have $l\ge l_{N,k}:=6N/(k+1)$. The actual value 
is slightly lower and is given by $l_{N,k}=(12N-9)/(2(k+1))$. 
This holds for $N\ge 2$, for $N=0$ there is nothing to do and the case 
$N=1$ can be reduced to the case $N=0$ by the expansion displayed in 
section \ref{s3}. This means that 
the polarisation identity must be invoked at least $l_{N,k}$ times.
The number of correction terms that are produced in each iteration 
triples the number of terms, hence the number of correction terms is of 
the order of $3^{l_{N,k}}$ and they are all of the same order of 
magnitude as the lowest order in $\hbar$ coefficient of (\ref{b.21}).\\
\\ 
What all of this means is of course that up to a controllable error of 
order $\hbar^{k+1}$ the value of (\ref{b.12}) is given by
\be \label{b.22}  
<\psi,p_1 \bar{f}_1 .. \bar{f}_N p_{N+1} \psi>
=<\psi,p_1 f^+_1 .. f^+_N p_{N+1} \psi>+O(\hbar^{k+1})
\ee
which is calculable and involves the fluctuations of $Q$ up to order 
$\hbar^k$ (we do not even have to calculate the constants 
$\beta_k$). In other 
words, we have shown that up to a controllable 
error of higher order in $\hbar$ it is allowed to use the power 
expansion of $f(t)=(1+t)^q$ up to order $2k+1$ in order to get a 
reliable value of (\ref{b.12}).

\subsection{Rigorous Estimate}
\label{s2.6}

\subsubsection{Step I: Proof of (\ref{b.18})}
\label{s2.6.1}

We want to prove (\ref{b.18}) and begin with the 
following observation\footnote{The factor $\frac{1}{<\psi,Q\psi>}$ is immaterial for the following power counting, because it can be factored out right fron the beginning. Therefore we will neglect it in the following discussion.}:\\
Since $(Q^2-<Q>^2)^{2(k+1)}=(Q-<Q>)^{2(k+1)} (Q+<Q>)^{2(k+1)}$ where 
$<Q>=<\psi,Q \psi>$ for our given coherent state $\psi$ we insert 
resolutions of unity for the matrix element of $(Q-<Q>)^{2(k+1)}$ 
between coherent states and can reduce our attention to the matrix 
element
\ba \label{b.23}
&& <\psi_1,[Q-<Q>]\psi_2> 
= i\epsilon_{j_1 j_2 j_3} 
[\prod_{I=1}^3 <\psi_{1I}, Y_I^{j_I} \psi'_{2I}>
-\prod_{I=1}^3 <\psi_{1I}, \psi'_{2I}> <\psi_I Y_I^{j_I} \psi_I>]
\\
&=& i\epsilon_{j_1 j_2 j_3} 
\left\{
[<\psi_{11}, Y_1^{j_1} \psi'_{21}>- <\psi_{11}, \psi'_{21}> <\psi_1 
Y_1^{j_1} \psi_1>] <\psi_{12}, Y_2^{j_2} \psi'_{22}>
<\psi_{13}, Y_3^{j_3} \psi'_{23}> \right.
\nonumber\\
&& +
<\psi_{11}, \psi'_{21}> <\psi_1 Y_1^{j_1} \psi_1> 
[<\psi_{12}, Y_2^{j_2} \psi'_{22}>-
<\psi_{12}, \psi'_{22}> <\psi_2 Y_2^{j_2} \psi_2>] 
<\psi_{13}, Y_3^{j_3} \psi'_{23}>
\nonumber\\
&& + 
<\psi_{11}, \psi'_{21}> <\psi_1 Y_1^{j_1} \psi_1> 
<\psi_{12}, \psi'_{22}> <\psi_2 Y_2^{j_2} \psi_2>
\times\nonumber\\ && \times \left.
[<\psi_{13}, Y_3^{j_3} \psi'_{23}>
-<\psi_{13}, \psi'_{23}> <\psi_3 Y_3^{j_3} \psi_3>]
\right\}
\nonumber
\ea
Here $\psi_1,\psi_2$ are any coherent states and the operators $Y^j_I$
for a given vertex $v$ are defined by 
$Y^j_I(v)=t [X^j_{I+}(v)-X^j_{I-}(v)]/2$, see (\ref{2.2}), where 
$t=\ell_P^2/a^2$ is the classicality parameter for some length scale $a$.
Here we have made use of the tensor product structure of the coherent 
states. It follows that 
\ba \label{b.24}
&& <\psi_{1I}, Y_I^{j_I} \psi'_{2I}>
-<\psi_{1I}, \psi'_{2I}> <\psi_I Y_I^{j_I} \psi_I>
\nonumber\\
& =& \frac{t}{2} \left\{
[<\psi_{1,I+},X^{j_I}_{I+} \psi_{2,I+}> 
- <\psi_{I+},X^{j_I}_{I+} \psi_{I+}> <\psi_{1,I+},\psi_{2,I+}>] 
<\psi_{1,I-},\psi_{2,I-}> \right. 
\nonumber\\
&& - \left.
<\psi_{1,I+},\psi_{2,I+}> 
[<\psi_{1,I-},X^{j_I}_{I-} \psi_{2,I-}> 
- <\psi_{I-},X^{j_I}_{I-} \psi_{I-}> <\psi_{1,I-},\psi_{2,I-}>]
\right\}
\ea
We now have reduced the computation to matrix elements of right 
invariant vector fields on a single copy of $SU(2)$ and can refer to
\cite{GCS} where we find 
\be \label{b.26}
it <\psi_{g_1},X^j \psi_{g_2}>=t <\psi_{g_1},\psi_{g_2}>
[\frac{2z_{12}}{t}+\frac{1}{z_{12}}-{\rm coth}(z_{12})]\frac{{\rm 
Tr}(g_1^\dagger 
g_2\tau_j)}{{\rm sh}(z_{12})}+O(t^\infty)
\ee
where 
$z_{12}$ is 
defined 
by ${\rm ch}(z_{12}):={\rm Tr}(g^\dagger g')/2$ for $g_1,g_2\in 
SL(2,\Cl)$.
Thus  
\ba \label{b.27}
&& it [<\psi_{g_1},X^j \psi_{g_2}>
-<\psi_{g_1},\psi_{g_2}>
<\psi_g,X^j \psi_g>
\nonumber\\
&=& <\psi_{g_1},\psi_{g_2}>
[\frac{z_{12}}{{\rm sh}(z_{12})} {\rm Tr}(g_1^\dagger g_2\tau_j)
-\frac{z}{{\rm sh}(z)} {\rm Tr}(g^\dagger g\tau_j) +O(t)]
\ea
where ${\rm ch}(z):={\rm Tr}(g^\dagger g)/2$. 

Thus, using resolutions of unity we have  
\ba \label{b.27a}
&& <\psi,(Q^2-<Q>^2)^{2(k+1)}\psi'>
= \int d\nu_1 .. d\nu_{2(k+1)}
<\psi,[Q-<Q>]\psi_1>\; <\psi_1,[Q-<Q>]\psi_2>\;.. 
\times\nonumber\\ && \times
<\psi_{2k+1},[Q-<Q>]\psi_{2(k+1)}>\; 
<\psi_{2(k+1)},[Q+<Q>]^{2(k+1)}\psi'>\; 
\ea
Now consider for $a=1,..,2k+1$ with $\psi_0:=\psi$ 
\ba \label{b.27b}
&& \int d\nu_a <\psi_{a-1},[Q-<Q>]\psi_a>\; <\psi_a,[Q-<Q>]\psi_{a+1}>
= \frac{1}{t^{18}} \prod_{I\sigma} \int_{\Rl^3} 
d^3p_{a,I\sigma} \int_{SU(2)} d\mu_H(U_{I\sigma})
\times\nonumber\\ && \times \exp(-[||p_{a,I\sigma}-p_{a-1,I\sigma}||^2 
+||p_{a,I\sigma}-p_{a+1,I\sigma}||^2
+||\theta_{a,I\sigma}-\theta_{a-1,I\sigma}||^2 
+||\theta_{a,I\sigma}-\theta_{a+1,I\sigma}||^2]/t)
\times \nonumber\\
&& \times
{\rm Pol}_4(z_{a,a-1,I\sigma},z_{a,a+1,I\sigma})
{\rm Pol}_2(z_{a,a-1,I\sigma})-2p_{I\sigma},
z_{a,a+1,I\sigma}-2p_{I\sigma})
\ea
Here we have explicitly written out the resolution measures 
$d\nu=cd^3p d\mu_H(U)[1+O(\hbar^\infty)]/t^3$ where $g=\exp(p_j\sigma_j) 
U,
U=\exp(-i\sigma_j \theta^j)$, $c$ is a positive constant of order unity
(essentially $\pi^{-3}$) 
and have used 
the form of the overlap function 
$<\psi_g,\psi_g'>
=\exp(-[||p-p'||^2+||\theta-\theta'||^2]/t)[1+O(\hbar^\infty)]$ 
close to $g'=g$ where it is sharply peaked. Moreover, close to 
$g=g'$ we have $z=p+p'+i(\theta'-\theta)$. The notation Pol$_n$ 
denotes 
a homogeneous polynomial of degree $n$.
In order to evaluate 
(\ref{b.27b}) we use translation invariance of the Haar measure 
$d\mu_H(U_{a,I\sigma})=
d\mu_H(U_{a,I\sigma} U_{a-1,I\sigma})^{-1}$ and notice that close to 
$\theta=0$ the Haar measure $d\mu_H=\sin^2(\theta) \sin(\theta) d\theta 
d\phi d\varphi$ with $\theta^2=\theta_j^2$, 
$\vec{\theta}/\theta=(\sin(\phi)\cos(\varphi),
\sin(\phi)\sin(\varphi),\cos(\phi))$ approaches the Lebesgue measure 
$d^3\theta$. Let us introduce $x^j_{a,I\sigma}:=(p^j_{a,I\sigma}-p^j)/s$ 
and $y^j_{a,I\sigma}=(\theta_{a,I\sigma}-\theta_{a-1,I\sigma})/s$ where
$s=\sqrt{t}$. Notice that $\theta/s\le \pi/s$ but we can estimate 
the integral from above by extending the integral over $\theta$ to 
infinity and in any case this just gives an error of higher order in 
$\hbar$. Then we can perform the Gaussian integrals and find to 
leading order in $t\propto \hbar$
\ba \label{b.27b1}
&& \int d\nu_a <\psi_{a-1},[Q-<Q>]\psi_a>\; <\psi_a,[Q-<Q>]\psi_{a+1}>
= t \prod_{I\sigma}  
\times\nonumber\\ && \times
\exp(-3[||p_{a+1,I\sigma}-p_{a-1,I\sigma}||^2 
+||\theta_{a+1,I\sigma}-\theta_{a-1,I\sigma}||^2]/(4t))
\times \nonumber\\
&& \times
{\rm Pol}'_4(p,sx_{a-1,I\sigma},s x_{a+1,I\sigma},
sy_{a-1,I\sigma},s y_{a+1,I\sigma})\;
{\rm Pol}'_2(x_{a-1,I\sigma},x_{a+1,I\sigma},
y_{a-1,I\sigma},y_{a+1,I\sigma})
\ea
where ${\rm Pol}'_n$ is a not necessarily homogeneous polynomial of 
degree $n$. 
Hence after carrying out all $2(k+1)$ integrals we arrive to leading 
order in $\hbar$ at an expression of the form 
\be \label{b.27c}
<\psi_1,(Q^2-<Q>^2)^{2(k+1)}\psi_2>=t^{k+1} <\psi_1,\psi_2>
{\rm Pol}'_{10(k+1)}(p) \; {\rm 
Pol}'_{2(k+1)}(p_1,p_2,\theta_1,\theta_2)
\ee
as claimed.

\subsubsection{Step II: Proof of (\ref{b.19})}
\label{s2.6.2}

In order to prove (\ref{b.19}) we must estimate 
$|<\psi_1, f \psi_2>|$. The method to estimate displayed below is not 
contained in \cite{GCS} so that we must be more detailed here. 
In fact, the estimate (\ref{b.19}) provides the {\it key part of 
the proof}, all the manipulations with Gaussian integrals that 
follow later are fairly standard.

Hence we need the explicit form of our coherent 
states \cite{GCS} which for a single copy of $SU(2)$ are given by
\be \label{b.28}
\psi_g(h)=\frac{\tilde{\psi}_g(h)}{||\tilde{\psi}_g(h)||},\;\;
\tilde{\psi}_g(h)=\sum_j d_j e^{-t j(j+1)/2} \chi_j(g h^{-1})
\ee
where the sum is over all spin quantum numbers $j=0,1/2,1,..$ while 
$d_j=2j+1$ and $\chi_j$ is the character of the $j-$th irreducible 
representation $\pi_j$ of $SU(2)$. Relevant for a given function $f$ of 
the volume operator located at a vertex $v$ are only the states 
$\psi_{g_{I\sigma}}$ with $I=1,2,3$ and $\sigma=\pm$ associated with the 
six outgoing edges adjacent to 
$v$. Here we use the notation of \cite{I,II}, that is, the edges of 
the algebraic graphs are labelled as $e_I(v)=:e_{I+}(v)$ denoting an edge 
outgoing 
from $v$ into the $I-$th direction. The other three edges are ingoing
but we can define the outgoing edges $e_{I-}(v)=[e_I(v-\hat{I})]^{-1}$. 
This has the consequence that 
\ba \label{b.29}
\tilde{\psi}_{e_{I-}(v)}(A)
&:= &\tilde{\psi}_{e_I(v-I)}(A)=\sum_j d_j e^{-t 
j(j+1)} [\pi_j(g_I(v-I))]_{mn} [\pi_j(A(e_I(v-I))^{-1})]_{nm}
\nonumber\\
&= &\sum_j \sqrt{d_j} e^{-t 
j(j+1)} [(\pi_j(g_I(v-I))^T]_{mn} T_{jmn}(A(e_{I-}(v)))
\nonumber\\
\tilde{\psi}_{e_{I+}(v)}(A)
&:= &\tilde{\psi}_{e_I(v)}(A)=\sum_j d_j e^{-t 
j(j+1)} [\pi_j(g_I(v))]_{mn} [\pi_j(A(e_I(v))^{-1})]_{nm}
\nonumber\\
&= &\sum_j \sqrt{d_j} e^{-t j(j+1)} [\pi_j(g_I(v))]_{mn} 
\overline{T_{jmn}(A(e_{I+}(v)))}
\ea
where $h\mapsto T_{jmn}(h)=\sqrt{2j+1} \pi_{jmn}(h)$ denotes a spin 
network function and we have used unitarity in the second step.
Let $g_{I+}:=g_I(v),\;g_{I-}:=[g_I(v-I)]^T$ then we obtain 
\ba \label{b.30}
&& <\psi_1,f\psi_2>\prod_{I,\sigma} 
||\tilde{\psi}_{1,I\sigma}||  ||\tilde{\psi}_{2,I\sigma}||  
\nonumber\\
&=& \sum_{\{j_{I\sigma}\}} \;
\prod_{I\sigma} [2j_{I\sigma}+1] e^{-tj_{I\sigma}(j_{I\sigma}+1)}
\sum_{|m_{I\sigma}|,|m'_{I\sigma}|,|n_{I\sigma}|,|n'_{I\sigma}|\le 
j_{I\sigma}} \times 
\nonumber\\
&& \times \prod_{I\sigma}
\overline{\pi_{j_{I\sigma} m_{I\sigma} n_{I\sigma}}(g_{1,I\sigma})}
\pi_{j_{I\sigma} m'_{I\sigma} n'_{I\sigma}}(g_{2,I\sigma})
\times \nonumber\\
&& \times 
<\otimes_I [\overline{T_{j_{I+},m_{I+},n_{I+}}}\otimes 
T_{j_{I-},m_{I-},n_{I-}}], 
f
\otimes_I [\overline{T_{j_{I+},m'_{I+},n'_{I+}}}\otimes 
T_{j_{I-},m'_{I-},n'_{I-}}]>
\nonumber\\
&=& \sum_{\{j_{I\sigma}\}} \;
\prod_{I\sigma} [2j_{I\sigma}+1] e^{-tj_{I\sigma}(j_{I\sigma}+1)}
\sum_{|m_{I\sigma}|,|m'_{I\sigma}|\le 
j_{I\sigma}} \times 
\nonumber\\
&& \times \prod_{I\sigma}
\pi_{j_{I\sigma} m'_{I\sigma} m_{I\sigma}}(g_{2,I\sigma} 
g_{1,I\sigma}^\dagger)
\times \nonumber\\
&& \times 
<\otimes_I [\overline{|j_{I+},m_{I+}>}\otimes 
|j_{I-},m_{I-}>], 
f
\otimes_I [\overline{|j_{I+},m'_{I+}>}\otimes 
|j_{I-},m'_{I-}>]>
\ea
where in the first step it was crucially exploited that the volume 
operator preserves the 
mutually orthogonal 
subspaces of the Hilbert space labelled by fixed spin quantum numbers.
In the second step we exploited that the spin network matrix element 
in the third line is non -- vanishing only when $n_{I\sigma}=n'_{I\sigma}$
and in that case is actually independent of $n_{I\sigma}$. The calculation 
then equals that in an abstract spin system so that we 
dropped the dependence on $n_{I\sigma}$ of the matrix element, thereby 
replacing $T_{jmn}$ by $|jm>$. The latter fact follows for instance 
from the detailed analysis in \cite{GT}.

Denote the factor in the fifth line of (\ref{b.30}) by 
$\overline{A_{\{m,m'\}}}$ and the factor in the sixth line by 
$B_{\{m,m'\}}$ then (\ref{b.30}) can be estimated by using the 
Schwarz inequality
\ba \label{b.31}
&& |<\psi_1,f\psi_2>|\prod_{I,\sigma} 
||\tilde{\psi}_{1,I\sigma}||\;\;  ||\tilde{\psi}_{2,I\sigma}||  
\nonumber\\
&\le& \sum_{\{j_{I\sigma}\}} \;
\prod_{I\sigma} [2j_{I\sigma}+1] e^{-tj_{I\sigma}(j_{I\sigma}+1)}
\;
\sqrt{\sum_{\{m,m'\}} |A_{m,m'}|^2}
\;\;\sqrt{\sum_{\{m,m'\}} |B_{m,m'}|^2}
\nonumber\\
&=& \sum_{\{j_{I\sigma}\}} \;
\prod_{I\sigma} [2j_{I\sigma}+1] e^{-tj_{I\sigma}(j_{I\sigma}+1)}
\;
\sqrt{\prod_{I\sigma} \chi_{j_{I\sigma}}(g_{1,I\sigma} 
g_{2,I\sigma}^\dagger g_{2,I\sigma} g_{1,I\sigma}^\dagger)}
\times \nonumber\\
&& \times 
\;\;\sqrt{\sum_{\{m\}} 
<\otimes_I [\overline{|j_{I+},m_{I+}>}\otimes 
|j_{I-},m_{I-}>], 
f
\otimes_I [\overline{|j_{I+},m_{I+}>}\otimes 
|j_{I-},m_{I-}]>}
\ea
where in the second step we used the completeness relation of the states 
$|jm>$ on the fixed $j$ subspace of the abstract spin system.\\
\\
We will now estimate the second square root in (\ref{b.31}).
Since certainly $f(t)=(1+t)^q-1\le 1+(1+t)-1=1+t$ for all $t\ge -1$ we 
have $f\le Q^2/<Q>^2$ and therefore have 
\be \label{b.32}
\sum_{\{m\}}
<\otimes_I [\overline{|j_{I+},m_{I+}>}\otimes 
|j_{I-},m_{I-}>], 
f
\otimes_I [\overline{|j_{I+},m_{I+}>}\otimes 
|j_{I-},m_{I-}]>
\le \sum_{\{m\}}\frac{1}{<Q>^2} 
||Q \otimes_{I\sigma} |j_{I\sigma},m_{I\sigma}>||^2 
\ee
where we exploited that in the angular momentum representation 
$\overline{|j,m>}=|j,-m>$ and that we sum over all $m$. Now we just have 
to evaluate the norm squared in (\ref{b.32}) by using elementary angular 
momentum calculus, we do not need to work in the recoupling basis 
\cite{Volume}.

Since the $Y_I^j,Y_J^k$ mutually commute for $I\not=J$ we have 
\ba \label{b.33}
Q^2 &=& 
-t^6 \epsilon^{IJK} \epsilon_{MNP} Y^1_I Y^2_J Y^3_K Y^1_M Y^2_N 
Y^3_P
\nonumber\\
&=&
- 3! t^6 Y^1_I Y^2_J Y^3_K Y^1_{[I} Y^2_J Y^3_{K]}
\nonumber\\
&=&
\sum_{I\not=J,K;\;J\not=K} \{
\nonumber\\
&& [Y^1_I Y^1_I]\;\;[Y^2_J Y^2_J]\;\; [Y^3_K Y^3_K]
-[Y^1_I Y^1_I]\;\;[Y^2_J Y^3_J]\;\; [Y^3_K Y^2_K]
\nonumber\\
&& +[Y^1_I Y^3_I]\;\;[Y^2_J Y^1_J]\;\; [Y^3_K Y^2_K]
-[Y^1_I Y^2_I]\;\;[Y^2_J Y^1_J]\;\; [Y^3_K Y^3_K]
\nonumber\\
&& +[Y^1_I Y^2_I]\;\;[Y^2_J Y^3_J]\;\; [Y^3_K Y^1_K]
-[Y^1_I Y^3_I]\;\;[Y^2_J Y^2_J]\;\; [Y^3_K Y^1_K]
\}
\ea
When we take the expectation value of (\ref{b.33}) with respect to
$\otimes_{I\sigma} |j_{I\sigma},m_{I\sigma}>$, only the first and fourth 
term survive because all other terms contain factors of the form
$Y^1_I Y^3_I,\; Y^3_I Y^1_I,\; Y^2_I Y^3_I,\; Y^3_I Y^2_I$
of which $Y^3_I$ is diagonal but $Y^1_I,\;Y^2_I$ is a linear combination 
of raising and lowering operators.

Since $iY^j_I=i(X^j_{I+}-X^j_{I-})/2$ is represented on 
the abstract spin system as $J^j_{I+}-J^j_{I-}$ where 
$J^j_{I\sigma}$ are the usual angular momentum operators we can now 
evaluate (\ref{b.33}) further
\ba \label{b.34}
&& ||Q  \otimes_{I\sigma} |j_{I\sigma},m_{I\sigma}>||^2
\nonumber\\
&=& \frac{t^6}{4}\sum_{I\not=J,K;\;J\not=K} 
<m_{I+},m_{I-}|[J_I^+]^2+[J_I^-]^2+J_I^+ J_I^- +J_I^- J_I^+|m_{I+},m_{I-}>
\times \nonumber\\ && \times
<m_{J+},m_{J-}|-[J_J^+]^2-[J_J^-]^2+J_J^+ J_J^- +J_J^- 
J_J^+|m_{J+},m_{J-}>
[m_{K+}-m_{K-}]^2
\nonumber\\
&& + \frac{t^6}{4}\sum_{I\not=J,K;\;J\not=K} 
<m_{I+},m_{I-}|[J_I^+]^2-[J_I^-]^2+[J_I^-, J_I^+]|m_{I+},m_{I-}>
\times \nonumber\\ && \times
<m_{J+},m_{J-}|[J_J^+]^2-[J_J^-]^2+[J_J^+, J_J^-] |m_{J+},m_{J-}>
[m_{K+}-m_{K-}]^2
\nonumber\\
&=& \frac{t^6}{16}\sum_{I\not=J,K;\;J\not=K} 
(m_{K+}-m_{K-})^2
\times\nonumber\\
&& \times
\{(j_{I+}(j_{I+}+1)+j_{I-}(j_{I-}+1)-m_{I+}^2-m_{I-}^2) 
(j_{J+}(j_{J+}+1)+j_{J-}(j_{J-}+1)-m_{J+}^2-m_{J-}^2) 
\nonumber\\
&& -(m_{I+}+m_{I-})(m_{J+}+m_{J-})\}
\ea
where $J^\pm_I=J^1_I\pm i J^2_I=J^\pm_{I+}-J^\pm_{I-}$ is a linear 
combination of the usual 
ladder operators and we 
have used the algebra 
$[J^j_{I\sigma},J^k_{J\sigma'}]=i\delta_{IJ}\delta_{\sigma\sigma'}
\epsilon_{jkl}  J_{I\sigma}^l$.

The second term in (\ref{b.34}) vanishes when summing over $\{m\}$ while
the first can be evaluated using the relation
\be \label{b.35}
\sum_{|m|\le j} m^2=\frac{j}{3}(j+1)(2j+1)
\ee
resulting in the compact formula 
\be \label{b.36}
\sum_{\{m\}} ||Q  \otimes_{I\sigma} |j_{I\sigma},m_{I\sigma}>||^2
=(\frac{2}{3})^5 t^6 [\prod_{I\sigma} (2j_{I\sigma}+1)]\;
[\prod_I [\sum_\sigma j_{I\sigma}(j_{I\sigma}+1)]]
\ee

As we wish to apply the Poisson resummation formula in order to evaluate 
the estimate (\ref{b.31}), we should estimate (\ref{b.36}) in terms of the 
integers $n_{I\sigma}=2j_{I\sigma}+1$. We have 
\ba \label{b.37}
&& (\frac{2}{3})^5 t^6 [\prod_{I\sigma} (2j_{I\sigma}+1)]\;
[\prod_I [\sum_\sigma j_{I\sigma}(j_{I\sigma}+1)]]
\nonumber\\ 
&=& 
(\frac{1}{3})^5 \frac{t^6}{2} [\prod_{I\sigma} n_{I\sigma} ]\;
[\prod_I [\sum_\sigma n_{I\sigma}^2 -1]]
\nonumber\\
&\le& 
(\frac{1}{3})^5 \frac{t^6}{2} \prod_I 
[n_{I+}^3 n_{I-} +n_{I-}^3 n_{I+}] 
\nonumber\\
&\le& 
(\frac{1}{3})^5 \frac{t^6}{16} \prod_I 
[n_{I+}+n_{I-}]^4
\ea
where we used $\sqrt{m m'}\le (m+m')/2$ and $m^2+(m')^2\le (m+m')^2$ for 
non negative integers $m,m'$.   

Next we evaluate the characters 
\be \label{3.38}
\chi_{j_{I\sigma}}(g_{1,I\sigma} g_{2,I\sigma}^\dagger 
g_{2,I\sigma} g_{1,I\sigma})=
\frac{{\rm sh}(n_{I\sigma} z_{I\sigma})}{{\rm sh}(z_{I\sigma})}
\ee
where 
\be \label{3.39}
{\rm ch}(z_{I\sigma}):=\frac{1}{2}{\rm Tr}(g_{1,I\sigma} 
g_{2,I\sigma}^\dagger g_{2,I\sigma} g_{1,I\sigma})
\ee
Since the argument of the trace in (\ref{3.39}) is a positive definite,
Hermitean, unimodular matrix, it is clear that $z_{I\sigma}$ is positive.
Since (\ref{3.38}) appears under a square root which would make the 
application of the Poisson resummation formula cumbersome, we use the 
elementary estimate 
\be \label{b.40}
\sqrt{2{\rm sh}(z)}\le 2{\rm ch}(z/2)
\ee
which holds for all $z\ge 0$. Putting everything together we find 
\ba \label{b.41}
&& |<\psi_1,f\psi_2>|\prod_{I,\sigma} 
||\tilde{\psi}_{1,I\sigma}||  ||\tilde{\psi}_{2,I\sigma}||  
\nonumber\\
&\le& 
\frac{t^3}{36 \sqrt{3}}
\sum_{\{n_{I\sigma}=1\}}^\infty \;
[\prod_{I\sigma}  
e^{-t(n_{I\sigma}^2-1)/4} \frac{{\rm ch}(n_{I\sigma} z_{I\sigma}/2)}{
\sqrt{{\rm sh}(z_{I\sigma})}}]\;\;
[\prod_I n_{I+} n_{I-} (n_{I+}+n_{I-})^2]
\nonumber\\
&\le& 
\frac{2t^3}{9\sqrt{3}}
\sum_{\{n_{I\sigma}=1\}}^\infty \;
[\prod_{I\sigma}  
e^{-t(n_{I\sigma}^2-1)/4} \frac{{\rm ch}(n_{I\sigma} z_{I\sigma}/2)}{
\sqrt{{\rm sh}(z_{I\sigma})}}]\;\;
[\prod_I (n_{I+}^4+n_{I-}^4)]
\nonumber\\
&=& 
\frac{2 t^3 e^{3t/2}}{9\sqrt{3}} \frac{1}{\sqrt{\prod_{I\sigma}
{\rm sh}(z_{I\sigma})}} \sum_{\sigma_1,\sigma_2,\sigma_3}
\times \nonumber\\
&&\times \prod_I \{
[\sum_{n_{I,\sigma_I}=1}^\infty \;n_{I,\sigma_I}^4 
 e^{-tn_{I,\sigma_I}^2/4} {\rm ch}(n_{I,\sigma_I} z_{I,\sigma_I}/2)]
\;\;[\sum_{n_{I,-\sigma_I}=1}^\infty  
 e^{-tn_{I,-\sigma_I}^2/4} {\rm ch}(n_{I,-\sigma_I} z_{I,-\sigma_I}/2)]
\}
\nonumber\\
&=& 
\frac{t^3 e^{3t/2}}{2^5\;9 \sqrt{3}} \frac{1}{\sqrt{\prod_{I\sigma}
{\rm sh}(z_{I\sigma})}} \sum_{\sigma_1,\sigma_2,\sigma_3}
\times \nonumber\\
&&\times \prod_I \{
[\sum_{n_{I,\sigma_I}=-\infty}^\infty \;n_{I,\sigma_I}^4 
 e^{-tn_{I,\sigma_I}^2/4} e^{n_{I,\sigma_I} z_{I,\sigma_I}/2}]
\;\;[-1+\sum_{n_{I,-\sigma_I}=-\infty}^\infty  
 e^{-tn_{I,-\sigma_I}^2/4} e^{n_{I,-\sigma_I} z_{I,-\sigma_I}/2}]
\} \nonumber\\
&\le&  
\frac{t^3 e^{3t/2}}{2^5 \;9\sqrt{3}} 
\frac{1}{\sqrt{\prod_{I\sigma}
{\rm sh}(z_{I\sigma})}} \sum_{\sigma_1,\sigma_2,\sigma_3}
\times \nonumber\\
&&\times \prod_I \{
[\sum_{n_{I,\sigma_I}=-\infty}^\infty \;n_{I,\sigma_I}^4 
 e^{-tn_{I,\sigma_I}^2/4} e^{n_{I,\sigma_I} z_{I,\sigma_I}/2}]
\;\;[\sum_{n_{I,-\sigma_I}=-\infty}^\infty  
 e^{-tn_{I,-\sigma_I}^2/4} e^{n_{I,-\sigma_I} z_{I,-\sigma_I}/2}]
\}
\ea
where we have used $2mn\le m^2+n^2$ in the second step.

Notice that 
\be \label{b.42}
||\psi_{j,I\sigma}||^2
=e^{t/4}\sum_{n=1}^\infty n e^{-tn^2/4} 
\frac{{\rm sh}(n y_{j,I\sigma})}{{\rm sh}(y_{j,I\sigma})} 
=\frac{e^{t/4}}{2{\rm sh}(y_{j,I\sigma})}
\sum_{n=-\infty}^\infty n e^{-tn^2/4} 
e^{n y_{j,I\sigma})}
\ee
where ${\rm ch}(y_{j,I\sigma}):={\rm Tr}(g_{j,I\sigma}^\dagger 
g_{j,I\sigma})/2,\;\;j=1,2$. 
Using the Poisson resummation formula \cite{GCS} we can now evaluate
\ba \label{b.43}
\sum_n n e^{-n^2 t/4} e^{yn}  &=& \frac{2\sqrt{\pi}}{s^3} 
e^{y^2/t} y
\nonumber\\
\sum_n e^{-n^2 t/4} e^{zn/2} &=& \frac{2\sqrt{\pi}}{s} e^{z^2/(4t)}\\
\nonumber\\
\sum_n n^4 e^{-n^2 t/4} e^{zn/2} &=& \frac{2\sqrt{\pi}}{s^9} 
e^{z^2/(4t)}(\frac{3}{2}s^4+\frac{3}{4} z^2 s^2+z^4/16)
\ea
where $s=\sqrt{t}$ and we have neglected $O(t^\infty)$ terms. We 
obtain the final formula
\ba \label{b.44}
|<\psi_1,f\psi_2>|  
&\le& 
\frac{1}{2^5\;9 \sqrt{3}} t^{-3}
\sqrt{\prod_{I\sigma} 
\frac{{\rm sh}(y_{1,I\sigma}){\rm sh}(y_{2,I\sigma})}
{y_{1,I\sigma} y_{2,I\sigma} {\rm sh}(z_{I\sigma})}} 
\prod_{I\sigma} 
e^{[z_{I\sigma}^2-2y_{1,I\sigma}^2-2y_{2,I\sigma}^2]/(4t)}
\times \nonumber\\
&&\times \ 
\sum_{\sigma_1,\sigma_2,\sigma_3}
\prod_I 
[\frac{3}{2}s^4+\frac{3}{4} z_{I\sigma_I}^2 s^2+z_{I\sigma_I}^4/16]
\ea
We want to show that (\ref{b.44}) is sharply peaked at 
$H_{1,I\sigma}=H_{2,I\sigma}$ where $g_{j,I\sigma}=U_{j,I\sigma} 
H_{j,I\sigma}$ is a polar decomposition into unitary and positive definite 
Hermitean matrices of the $SL(2,\Cl)$ matrices. Parametrising
$H=\exp(p_j \sigma_j)={\rm ch}(p)+\sigma_j p_j {\rm sh}(p)/p$ where
$p^2=p_j p_j$ and $\sigma_j$ are the Pauli matrices we easily find 
$y_{j,I\sigma}=2p_{j,I\sigma}$ and  
\be \label{b.45}
{\rm ch}(z_{I\sigma})=
{\rm ch}(y_{1,I\sigma}) {\rm ch}(y_{2,I\sigma})
+ c_{I\sigma} 
{\rm sh}(y_{1,I\sigma}) {\rm sh}(y_{2,I\sigma}),\;\;
c_{I\sigma}=\frac{p^j_{1,I\sigma}p^j_{2,I\sigma}}{p_{1,I\sigma}p_{2,I\sigma}}
\ee
Since $c_{I\sigma}\in [-1,1]$ we find from basic hyperbolic identities
\be \label{b.46} 
|y_{1,I\sigma}-y_{2,I\sigma}|\le z_{I\sigma} \le
(y_{1,I\sigma}+y_{2,I\sigma})
\ee
Hence the argument of the exponent in (\ref{b.44}) satisfies the following 
relation
\be \label{b.47}
z_{I\sigma}^2-2y_{1,I\sigma}^2-2y_{2,I\sigma}^2
\le -(y_{1,I\sigma}-y_{2,I\sigma})^2 \le 0
\ee
where equality is reached if and only if $c_{I\sigma}=1$ and 
$y_{1,I\sigma}=y_{2,I\sigma}$, that is, $H_{1,I\sigma}=H_{2,I\sigma}$ as 
claimed. This property of (\ref{b.44}) will be crucial for what 
follows, because otherwise in the resolutions of unity that will follow 
below, if we would not have the Gaussian decay just established, the 
corresponding integrals would blow up.

\subsubsection{Step III: Proof of (\ref{b.16})}

Next we prove (\ref{b.16}).
From \cite{GCS} we recall that the measure $\nu$ involved in the 
resolution of unity is given, up to a $O(t^\infty)$ correction by 
\be \label{b.48}
d\nu(g)=\frac{c}{t^3} d^3p d\mu_H(U),\;\;g=\exp(p^j\sigma_j)\; U
\ee
where $c$ is a numerical constant of order unity. Hence the resolution 
measure is up to the important factor $1/t^3$ essentially given by the 
Liouville measure on $T^\ast(SU(2))$. Also recall \cite{GCS} that the 
overlap function is essentially (i.e. up to $O(t^\infty)$ 
corrections) given by 
\ba \label{b.49}
|<\psi_g,\psi_{g'}>|^2 &=& \frac{|\tilde{z}| {\rm sh}(p) {\rm 
sh}(p')}{|{\rm 
sh}(\tilde{z})| 
pp'} e^{-\frac{4\Delta^2+2\tilde{\theta}^2+2\delta^2}{t}}
\nonumber\\
\Delta^2 &=& p^2+(p')^2-\tilde{p}^2/2
\nonumber\\
{\rm ch}(2\tilde{p}) &=& 
(1+c) {\rm ch}^2(p+p')
+(1-c) {\rm ch}^2(p-p')-1,\;c=\frac{p_j p'_j}{pp'}
\nonumber\\
{\rm ch}(\tilde{z}) &=& {\rm ch}(\tilde{p})\cos(\tilde{\theta})+i{\rm 
sh}(\tilde{p}) \sin(\tilde{\theta}) \cos(\tilde{\alpha}),\;
\cos(\tilde{\alpha})=\frac{\tilde{p}_j 
\tilde{\theta}_j}{\tilde{p}\tilde{\theta}} 
\nonumber\\
\tilde{\delta}^2 &=& 
\tilde{p}^2-\tilde{s}^2+\tilde{\phi}^2-\tilde{\theta}^2,\;\;
\tilde{z}=\tilde{s}+i\tilde{\phi}
\ea 
where the notation is as follows: We set 
$g=\exp(p_j\sigma_j)U,\;g'=\exp(p'_j\sigma_j)U'$ and 
$\exp(p_j\sigma_j) \;\exp(p'_j\sigma_j)=\exp(\tilde{p}_j\sigma_j)
U^{\prime\prime}$ and finally $\tilde{U}=U^{-1} U' (U^{\prime\prime})^{-1}
=\exp(-i\tilde{\theta}_j\sigma_j)$. Here 
$\tilde{\theta}^2=\tilde{\theta}_j\tilde{\theta}_j$ and all angles 
$\beta,\tilde{\alpha},\tilde{\phi}$ lie in $[0,\pi]$. All quantities 
are uniquely determined by the above formulas. Moreover one can show:\\
1.\\
$\Delta^2\ge 0$ where equality is reached if and only if $p_j=p'_j$.\\
2.\\ 
$\delta^2\ge 0$ where equality is reached if and only if either
a) $\tilde{\phi}=\tilde{\theta}, |s|=\tilde{p}, |\cos(\tilde{\alpha})|=1$
or b) $s=\tilde{p}=0$ or c) $\tilde{\phi}=\tilde{\theta}=0,\pi$.\\
Since $\delta^2\ge 0$, in order that the exponent in the first line of 
(\ref{b.49}) vanishes we must have at least $\Delta^2=\tilde{\theta}^2=0$.
Since $\Delta^2=0$ is equivalent with $p_j=p'_j$ this means that 
$U^{\prime\prime}=1$ and therefore $\tilde{\theta}=0$ means $U=U'$, that 
is, $g=g'$ altogether. Clearly then $z=\tilde{p}=2p$, hence 
$\tilde{\phi}=\tilde{\theta}=0$ so we are in case c) and thus 
$\delta^2=0$. 

What all of this means is that the overlap function 
essentially takes the form of a Gaussian peaked at $g=g'$. The more 
complicated structure is due to the fact that we are dealing with a non -- 
Abelean group rather than a vector space. Since the overlap function 
vanishes rapidly when $g\not=g'$, let us expand (\ref{b.49}) in terms of 
$\Delta p_j=p'_j-p_j$ and $\Delta\theta_j=\theta'_j-\theta_j$ where 
$U=\exp(-i\theta_j\sigma_j),\;U'=\exp(-i\theta'_j\sigma_j)$.
Then it is not difficult to see that 
\ba \label{b.49a}
|<\psi_g,\psi_{g'}>|^2 &=& 
e^{-\frac{[\Delta p]^2+[\Delta\theta]^2}{t}}
\nonumber\\
|<\psi_1,f\psi_2>|  
&\le& 
\frac{a^6 2^7}{9 \sqrt{3}} t^{-3}
e^{-\sum_{I,\sigma} \frac{[\Delta p_{I\sigma}]^2}{t}}
\; [\prod_I p_{I+}^4+p_{I-}^4]
\ea
where $\Delta p_{I\sigma}^j=p_{1,I\sigma}^j-p_{2,I\sigma}^j$.\\
\\
A further result that we need from \cite{GCS} is the following:\\
The polynomials $P$ are 
typically just (products) holonomy operators in the spin $1/2$ 
representation for which holds 
\ba \label{b.49b}
<\psi_{g}, h_{AB} \psi_{g'}> &=& <\psi_g,\psi_{g'}>
\left\{
[g'_{AB} {\rm ch}(z/2)
+(g'\tau_j)_{AB} \frac{{\rm Tr}(g^\dagger g' \tau_j)}{2 {\rm sh}(z)}
{\rm sh}(z/2)] \right.
\\
&& \left.
+t[\frac{{\rm sh}(z/2)}{2z}+({\rm ch}(z)g'-g^\dagger 
(g')^2)_{AB} \frac{{\rm coth}(z) {\rm sh}(z/2)-\frac{1}{2}
{\rm ch}(z/2)}{z{\rm sh}(z)}] +[O(t^\infty)]
\right\}
\nonumber
\ea
Here $g^\dagger g'={\rm ch}(z)-i\tau_j z_j/z {\rm sh}(z)$. Due to the 
Gaussian prefactor $<\psi_g,\psi_{g'}>$ the first term of 
order $O(t^0)$ contributes only at $g=g'$ and there is given by $U_{AB}$
where $g=HU,\;g'=H'U'$ are the polar decompositions of $g,g'$ with
$H=\exp(ip_j \tau_j),\;U=\exp(\theta_j \tau_j)$. At 
$g\not=g'$ this term as well as the one of order $t^1$ and $t^\infty$ 
are exponentially bounded in $p,p'$ which is suppressed by the Gaussian.
More precisely, in the vicinity of $g'=g$ where the main contribution to 
the integral that we are interested in comes from, the term of zeroth 
order (in 
$t$) in (\ref{b.16}) becomes approximately 
\be \label{b.49c}
<\psi_{g}, h_{AB} \psi_{g'}> = 
e^{-\frac{1}{t}[(p'_j-p_j)^2+(\theta'_j-\theta_j)^2]}
\; \exp(i(p'_j-p_j)\tau_j/2) \; \exp((\theta'_j+\theta)_j \tau_j/2)
\ee
as claimed.

\subsubsection{Step IV: Performance of the Resolution Integrals}
\label{s2.6.4}

The way to extract the leading $\hbar$ power of the correction terms is 
now as follows: Recalling (\ref{b.15}), the final integrals to be 
computed are of the form
\ba \label{b.50}
&& \int d\nu_1 .. \int d\nu_{N/2}\;\;
\int d\nu'_1 .. \int d\nu'_{N/2}\;\;
\int d\tilde{\nu}_1 .. \int d\tilde{\nu}_l \;\;
\int d\tilde{\nu}'_1 .. \int d\tilde{\nu}'_l 
\nonumber\\
&& \int d\nu_{00} .. \int d\nu_{0 N/2}\;\;
\int d\nu'_{00} .. \int d\nu'_{0 N/2}\;\;
\int d\tilde{\nu}_{01} .. \int d\tilde{\nu}_{0l} \;\;
\int d\tilde{\nu}'_{01} .. \int d\tilde{\nu}'_{0l} 
\times \nonumber\\
&& \times
|<\psi, P_1 \psi_{00}>|\;
|<\psi_{00},f_1 \psi_1>|\; 
|<\psi_1, P_2 \psi_{01}>|\;
|<\psi_{01}, f_2 \psi_2>| \;..
\times\nonumber\\ && \times
|<\psi_{0N/2-1} f_{N/2} \psi_{N/2}>|\;
|<\psi_{N/2}, P_{N/2+1} \psi_{0N/2}>|\;
\times \nonumber\\
&& \times
|<\psi, P_1 \tilde{\psi}_{00}>|\;
|<\tilde{\psi}_{00},f_1 \tilde{\psi}_1>|\; 
|<\tilde{\psi}_1, P_2 \tilde{\psi}_{01}>|\;
|<\tilde{\psi}_{01}, f_2 \tilde{\psi}_2>| 
\;..\;
\times \nonumber\\ && \times
|<\tilde{\psi}_{N/2-1} f_{N/2} \tilde{\psi}_{N/2}>|\;
|<\tilde{\psi}_{N/2}, P_{N/2+1} \tilde{\psi}_{0N/2}>|\;
\times \nonumber\\
&& \times |<\psi_{0N/2},[\Delta f'_1] \psi'_1>|\;
|<\psi'_1, P'_1 \psi'_{01}>|\;
|<\psi'_{01},[\Delta f'_2]\psi'_2>|\; 
|<\psi'_2, P'_2 \psi'_{02}>|\;..\;
\times \nonumber\\ && \times
|<\psi'_{0l-1},[\Delta f'_l]\psi'_l>|
\times \nonumber\\
&& \times 
|<\tilde{\psi}_{0N/2},[\Delta f'_1] \tilde{\psi}'_1>|\;
|<\tilde{\psi}'_1, P'_1 \tilde{\psi}'_{01}>|\;
|<\tilde{\psi}'_{01},[\Delta f'_2] \tilde{\psi}'_2>|\; 
|<\tilde{\psi}'_2, P'_2 \tilde{\psi}'_{02}>|\;..\;
\times \nonumber\\ && \times
|<\tilde{\psi}'_{0l-1},[\Delta f'_l]\tilde{\psi}'_l>|\;
\times \nonumber\\ && \times 
|<\psi'_l,P'_l\psi'_{0l}>|\;
|<\tilde{\psi}'_l,P'_l \tilde{\psi}'_{0l}>|\;
|<\psi'_{0l},[\Delta f'_{l+1}]\tilde{\psi}'_{0l}>|
\ea 
As we have seen, factors of the form $|<\psi_1,f\psi_2>|$ are 
to leading order in $\hbar$ essentially estimated by
products of Gaussians in $p_{2,I\sigma}^j-p_{2,I\sigma}^j$ times a 
homogeneous polynomial in $p^j_{1,I\sigma},\;p_{2,I\sigma}$ of order 
twelve. On the other hand,
factors of the form $<\psi_1,[\Delta F]\psi_2>$ could be written to 
leading order in $\hbar$ as 
Gaussians in $p^j_{2,I\sigma}-p^j_{1,I\sigma}$ and 
$\theta^j_{2,I\sigma}-\theta^j_{1,I\sigma}$ times a homogeneous polynomial 
of order $5(2k+2)$ in $z^j_{I\sigma}$ times a homogeneous polynomial of 
order
$2k+2$ in $z^j_{I\sigma}-p^j_{I\sigma}$ 
where for $g_{1,I\sigma}\approx g_{2,I\sigma}$ we have   
\be \label{b.51}
z^j_{I\sigma}\approx [p_{2,I\sigma}+p_{1,I\sigma}]+i
[\theta_{2,I\sigma}-\theta_{1,I\sigma}]
\ee
as follows from the explicit formulas proved in \cite{GCS}.
Here $p^j_{I\sigma}=i<\psi,X^j_{I\sigma}\psi>$ is the expectation value 
of the right invariant vector fields with respect to the state $\psi$ 
under consideration. Finally, factors of the form 
$<\psi_1, P \psi_2>$ could be written, to leading order in $\hbar$  as 
Gaussians in $p^j_{2,I\sigma}-p^j_{1,I\sigma}$ and 
$\theta^j_{2,I\sigma}-\theta^j_{1,I\sigma}$ times sums of products 
of matrix 
elements of the $SL(2,\Cl)$ elements 
$\exp(-i (p^j_{2,I\sigma}-p^j_{1,I\sigma})\tau_j/2)$,
$\exp(\theta^j_{2,I\sigma}+\theta^j_{1,I\sigma})\tau_j/2)$.
  
We now carry out the various integrals to leading order in $\hbar$. 
We will drop the indices $I,\sigma,j$ in the following power counting 
argument, hence for instance $d\nu_a$ means an integral over all the 
$p^j_{a,I\sigma},\theta^j_{a,I\sigma}$ and by $p_a$ we mean the 18
variables $p^j_{a,I\sigma}$ etc. and correspondingly
$||p_a||^2:=\sum_{I\sigma j} (p^j_{a,I\sigma})^2$. Further
$dp_a:=\prod_{I\sigma j} dp^j_{a,I\sigma},\;
d\mu_H(U_a):=\prod_{I\sigma} d\mu_H(U_{a,I\sigma})$. 
Also, for simplicity we assume 
that, as it happens in our application, all volume 
operators are with respect to same vertex $v$, otherwise we just have to 
introduce more notation which however does not change the argument.
Also for simplicity we assume that we just have to consider holonomies
along the edges $e_{I\sigma}(v)$, for loops we would otherwise just have 
to use more resolution integrals but they do not change the $\hbar$ 
power. 

The first type of integral we consider is (for $a=0,..,N/2-1$ with 
$\psi_0:=\psi$)
\ba \label{b.51a}
&&\int d\nu_{0a} |<\psi_a, P_{a+1} \psi_{0a}>| \; |<\psi_{0a}, F_{a+1} 
\psi_{a+1}>|
\nonumber\\
&\le & \frac{1}{t^3} \int \frac{dp_{0a} d\mu_H(U_{0a})}{t^{18}}
\exp(-[||p_{0a}-p_a||^2+||p_{0a}-p_{a+1}||^2+||\theta_{0a}-\theta_a||^2]/t)
\times \nonumber\\ \times &&
|{\rm Pol}_{12}(p_{0a},p_{a+1}) \;\;{\rm Pol}_1(\exp([p_{0a}-p_a]/2),
\exp(i[\theta_{0a}+\theta_a]/2)|
\ea
where Pol$_n$ denotes a homogeneous polynomial of degree $n$ of 
the variables indicated and we have 
only kept the leading order in $\hbar$. Here $\exp([p_{0a}-p_a]/2)$ stands 
for the collection of group elements 
$\exp(\sigma_j (p^j_{0a,I\sigma}-p^j_{a,I\sigma})/2)$ etc.

Using 
translation invariance of the Haar measure, we get 
$d\mu_H(U_{0a})=d\mu_H(U_{0a} U_a^{-1})$ and since the Gaussian in 
(\ref{b.51}) receives its essential support at $U_{0a}=U_a$ the Haar 
measure can be replaced by the Lebesgue measure, that is, for small
$\theta^j$ the Haar measure  
$d\mu_H(U)=\sin^2(\theta) \sin(\phi) d\theta d\phi d\varphi$ with
$\vec{\theta}/\theta=
(\sin(\phi)\cos(\varphi),\sin(\phi)\sin(\varphi),\cos(\phi))$ approaches 
$d^3\theta$ with integration domain a ball of radius $\pi$. The polynomial 
involving $\exp((\theta_{0a}+\theta_a)/2)$ can be estimated from above by 
a polynomial in $\exp((p_{0a}-p_a)/2)$ alone because matrix elements 
of $SU(2)$ group elements are bounded by $1$. 

Let us introduce $x_{0a}:=(p_{0a}-p)/s,\;
x_a:=(p_a-p)/s,\;y_{0a}:=(\theta_{0a}-\theta_a)/s$ where $s=\sqrt{t}$
then (\ref{b.51}) can be further estimated by 
\ba \label{b.52}
&&\int d\nu_{0a} |<\psi_a, P_{a+1} \psi_{0a}>| \; |<\psi_{0a}, f_{a+1} 
\psi_{a+1}>|
\nonumber\\
&\le & \frac{1}{t^{21}} \int_{\Rl^{18}} dx_{0a} \int_{||y_{0a}||\le \pi} 
dy_{0a} 
\exp(-[||x_{0a}-x_a||^2+||x_{0a}-x_{a+1}||^2+||y_{0a}||^2])
\times\nonumber\\ && \times
|{\rm Pol}_{12}(p+sx_{0a},p+s x_{a+1}) 
\;\;{\rm Pol}_1(\exp(s[x_{0a}-x_a]),
\exp(\theta_{0a}+\theta_a))|
\nonumber\\
&\le & \frac{1}{t^{21}} \int_{\Rl^{18}} dx_{0a} \int_{||y_{0a}\le 
\pi/s} 
dy_{0a} 
\exp(-[||x_{0a}-x_a||^2+||x_{0a}-x_{a+1}||^2+||y_{0a}||^2])
\times\nonumber\\ && \times
|{\rm Pol}_{12}(p+sx_{0a},p+s x_{a+1}) 
\;\;{\rm Pol}_1(\exp(s[x_{0a}-x_a]))|
\nonumber\\
&=& \frac{\pi^{18}}{t^3} {\rm Pol}_{12}(p) \exp(-3(||x_a - 
x_{a+1}||+s)^2/4)
\ea
where we have used basic properties of Gaussian integrals, used
$\int_0^{\pi/s} d\theta I \le \int_0^\infty d\theta I$ for positive 
integrand $I$ and dropped subleading orders of $s$. We will also drop the 
factor $\exp(-3s ||x_a-x_{a+1}||)$ in (\ref{b.52}) in what follows as it 
just leads to 
higher order corrections. 

We can now perform the integrals corresponding to the measures $\nu_1,..,
\nu_{N/2-1}$. Since the integrand no longer depends on $U_1,..,U_{N/2-1}$ 
the Haar measure part of those measures just integrates to one and we are 
left with an integral of the form 
\ba \label{b.53}
&& \frac{c}{t^{3N/2}} 
\frac{1}{s^{18(N/2-1)}}
{\rm Pol}_{6N}(p) \int dx_1 .. \int dx_{N/2-1} 
\;\;\exp(-3||x-x_1||^2/4) 
.. 
\exp(-3||x_{N/2-1}-x_{N/2}||^2)   
\nonumber\\
&=&  \frac{c'}{t^{3N/2}} \frac{1}{s^{18(N/2-1)}} {\rm Pol}_{6N}(p) \exp(-3||x-x_{N/2}||^2/(2N))
\ea
where $c,c'$ are numerical constants and we have made use of the Markov 
property $\int dy K_s(x,y) K_t(y,z)=c K_{s+t}(x,z)$ with 
$K_t(x,y)=\exp(-(x-y)^2/t$. The negative power of $s=\sqrt{t}$ comes from 
the fact that $d\nu=dp d\mu_H(U)/t^{18}$ but only $1/s^{18}$ was 
absorbed by the Gaussian integral in $p$.

Next we perform the integral corresponding to $\nu_{0N/2}$ which is of the 
form 
\ba \label{b.54}
&& \int d\nu_{0N/2} 
|<\psi_{N/2},P_{N/2+1} \psi_{0 N/2}>| \;
|<\psi_{0 N/2},[\Delta F'_1]  \psi'_1>|
\le c \int \frac{dp_{0 N/2} d\mu_H(U_{0 N/2})}{t^{18}} 
\times \nonumber\\  && \times   
\exp(-[||p_{0 N/2}-p_{N/2}||^2+||p_{0 N/2}-p'_1||^2+ 
[||\theta_{0 N/2}-\theta_{N/2}||^2+||\theta_{0 N/2}-\theta'_1||^2]/t)
\times\nonumber\\
&& \times
| {\rm Pol}_1(\exp((p_{0 N/2}-p_{N/2})/2),\exp(i(\theta_{0 
N/2}-\theta_{N/2})/2)\;\;
{\rm Pol}_{10(k+1)}(p_{0 N/2}+p'_1+i(\theta'_1-\theta_{0 N/2}) 
\times\nonumber\\ && \times
{\rm Pol}_{2(k+1)}(p_{0 N/2}+p'_1-2p+i(\theta'_1-\theta_{0 N/2})| 
\ea
We introduce new integration variables 
$x_{0 N/2}=(p_{0 N/2}-p)/s,\;
x'_1=(p'_1-p)/s,\;y_{0 N/2}=(\theta_{0 N/2}-\theta'_1)/s$, replace the 
Haar measure by the Lebesgue measure after an appropriate shift in 
$SU(2)$ as above, extend the integration domain from a ball to all 
of $\Rl^3$, estimate entries of $SU(2)$ group elements from above by 
$1$ 
and can estimate (\ref{b.54}) further by 
\ba \label{b.55}
&& \int d\nu_{0N/2} 
|<\psi_{N/2},P_{N/2+1} \psi_{0 N/2}>| \;
|<\psi_{0 N/2},[\Delta F'_1]  \psi'_1>|
\le  \int dx_{0 N/2} dy_{0 N/2}) 
\times\nonumber\\ &&  \times
\exp(-[||x_{0 N/2}-x_{N/2}||^2+||x_{0 N/2}-x'_1||^2+ 
[||y_{0 N/2}+(\theta'_1-\theta_{N/2})/s||^2+||y_{0 N/2}||^2])
\times\nonumber\\
&& \times
| {\rm Pol}_1(\exp(s(x_{0 N/2}-x_{N/2})/2)
{\rm Pol}_{10(k+1)}(2p +s[x_{0 N/2}+x'_1-iy_{0 N/2}]) 
\times\nonumber\\ &&\times
{\rm Pol}_{2(k+1)}(s[x_{0 N/2}+x'_1-iy_{0 N/2}])| 
\nonumber\\
&=& c t^{k+1} {\rm Pol}_{10(k+1)}(p)\; {\rm Pol}'_{2(k+1)}(x'_1)
\exp([-3[||x'_1-x_{N/2}||^2+||\theta'_1-\theta_{N/2}||^2/t]/4)
\ea
where we again used basic properties of Gaussian integrals and dropped 
subleading orders in $\hbar$. We used the notation Pol$'_n$ for a generic 
inhomogeneous polynomial of degree $n$ and performed estimates of the 
form $|P_{2(k+1)}(x+iy)|=P_{k+1}(x^2+y^2)$.  

Combining (\ref{b.53}) and (\ref{b.55}) we can perform the integral over
$\nu_{N/2}$ using the same manipulations and end up with an error estimate 
so far given by 
\be \label{b.56}
= c t^{k+1-3N/2} \frac{1}{s^{18(N/2-1)}} {\rm Pol}_{6N+10(k+1)}(p) 
{\rm Pol}'_{2(k+1)}(x'_1) \exp(-3||x-x'_1||^2/(2N+4))
\ee
where $x=p/s$ and $p$ corresponds to the external state $\psi$.
Next consider the integral for ($a=1,..,l-1$)
\be \label{b.57} 
\int d\nu'_{0a} |<\psi'_a, P'_a \psi'_{0a}>| \;|<\psi'_{0a},[\Delta 
F'_{a+1} \psi'_{a+1}>|
\ee
which can be estimated just like (\ref{b.54}) with the obvious changes in 
the integration variables, resulting in 
\ba \label{b.58} 
&& \int d\nu'_{0a} |<\psi'_a, P'_a \psi'_{0a}>| \;|<\psi'_{0a},[\Delta 
F'_{a+1} \psi'_{a+1}>|
\nonumber\\
& \le & 
c t^{k+1} {\rm Pol}_{10(k+1)}(p) {\rm Pol}'_{2(k+1)}(x'_{a+1})
\exp([-3[||x'_{a+1}-x'_a||^2+||\theta'_{a+1}-\theta'_a||^2/t]/4)
\ea
We can now combine (\ref{b.56}) and (\ref{b.58}) and perform the integrals
corresponding to the measures $\nu'_1,..,\nu'_{l-1}$. We use 
translation invariance of $d\mu_H(U'_a)=d\mu_H(U'_a 
(U'_{a+1})^{-1})$ and introduce $y'_a=(\theta'_a-\theta'_{a+1})/s$ and 
arrive at the estimate so far 
\be \label{b.59}
c t^{l(k+1)-3N/2} \frac{1}{s^{18(N/2-1)}} {\rm Pol}_{6N+10l(k+1)}(p) 
{\rm Pol}'_{2(k+1)}(x'_l) \exp(-3||x-x'_l||^2/(2N+4l))
\ee
where we absorbed the constants corresponding to Gaussian integrals over
polynomials into $c$.

All we did so far can also be done to the corresponding integrals 
involving the measures with a tilde resulting in the estimate
\be \label{b.60}
c t^{l(k+1)-3N/2} \frac{1}{s^{18(N/2-1)}} {\rm Pol}_{6N+10l(k+1)}(p) 
{\rm Pol}'_{2(k+1)}(\tilde{x}'_l) \exp(-3||x-\tilde{x}'_l||^2/(2N+4l))
\ee
It remains to perform the integral
\ba \label{b.61}
&& c t^{2l(k+1)-3N} \frac{1}{t^{18(N/2-1)}} {\rm Pol}_{12N+20l(k+1)}(p) 
\int d\nu'_l \int d\tilde{\nu}'_l  \int d\nu'_{0l} 
\int d\tilde{\nu}'_{0l}  
\times \nonumber\\
&& \times 
{\rm Pol}'_{2(k+1)}(x'_l) \exp(-3||x-x'_l||^2/(2N+4l))\;
{\rm Pol}'_{2(k+1)}(\tilde{x}'_l) \exp(-3||x-\tilde{x}'_l||^2/(2N+4l))
\times \nonumber\\
&& \times
|<\psi'_l,P'_l \psi'_{0l}>|\;
|<\tilde{\psi}'_l,P'_l \tilde{\psi}'_{0l}>|\;
|<\psi'_{0l}, [\Delta F'_{l+1}] \tilde{\psi}'_{0l}>|
\ea
We perform first the integral corresponding to $\psi'_{0l}$ using the same 
manipulations as in (\ref{b.55}) leading to an estimate of (\ref{b.61})
given by 
\ba \label{b.62}
&& c t^{(2l+1)(k+1)-3N} \frac{1}{t^{18(N/2-1)}} 
{\rm Pol}_{12N+10(2l+1)(k+1)}(p) 
\int d\nu'_l \int d\tilde{\nu}'_l  
\int d\tilde{\nu}'_{0l}  
\times \nonumber\\
&& \times 
{\rm Pol}'_{2(k+1)}(x'_l) \exp(-3||x-x'_l||^2/(2N+4l))\;
{\rm Pol}'_{2(k+1)}(\tilde{x}'_l) \exp(-3||x-\tilde{x}'_l||^2/(2N+4l))
\times \nonumber\\
&& \times
{\rm Pol}'_{2(k+1)}(x'_{l+1})\;
\exp([-3[||x'_l-\tilde{x}'_{0l}||^2+
||\theta'_l-\tilde{\theta}'_{0l}||^2/t]/4)\;
|<\tilde{\psi}'_l,P'_l \tilde{\psi}'_{0l}>|\;
\nonumber\\
&\le&
c t^{(2l+1)(k+1)-3N} \frac{1}{t^{18(N/2-1)}} 
{\rm Pol}_{12N+10(2l+1)(k+1)}(p) 
\int d\nu'_l \int d\tilde{\nu}'_l  
\int d\tilde{\nu}'_{0l}  
\times \nonumber\\
&& \times 
{\rm Pol}'_{2(k+1)}(x'_l) \exp(-3||x-x'_l||^2/(2N+4l))\;
{\rm Pol}'_{2(k+1)}(\tilde{x}'_l) \exp(-3||x-\tilde{x}'_l||^2/(2N+4l))
\times \nonumber\\
&& \times
{\rm Pol}'_{2(k+1)}(x'_{l+1})\;
\exp([-3[||x'_l-\tilde{x}'_{0l}||^2+
||\theta'_l-\tilde{\theta}'_{0l}||^2/t]/4)\;
\exp([-[||\tilde{x}'_l-\tilde{x}'_{0l}||^2+
||\tilde{\theta}'_l-\tilde{\theta}'_{0l}||^2/t])
\times \nonumber\\
&& \times 
{\rm Pol}_1(\exp((\tilde{p}'_l-\tilde{p}'_{0l})/2),
\exp(i[\tilde{\theta}'_l+\tilde{\theta}'_{0l}]/2)
\ea
Using translation invariance 
$d\mu_H(U'_l)=d\mu_H(U'_l (\tilde{U}'_{0l})^{-1})$ and 
$d\mu_H(\tilde{U}'_l)=d\mu_H(\tilde{U}'_l (\tilde{U}'_{0l})^{-1})$ 
we can perform the Gaussians in  
$\theta'_l-\tilde{\theta}'_{0l},\;
\tilde{\theta}'_l-\tilde{\theta}'_{0l}$. After this, nothing depends on 
$\tilde{\theta}'_{0l}$ any more and the corresponding Haar measure 
integrates to unity. Finally the remaining three momentum Gaussian 
integrals can 
also be performed and give a constant of order unity times $s^{-18}$ 
from 
one missing Gaussian factor in the angle variables. Thus the final 
estimate is given by
\be \label{b.63}
c t^{(2l+1)(k+1)-3N-18(N/2-1/2)} 
{\rm Pol}_{12N+10(2l+1)(k+1)}(p) 
\ee
The constant $c$ is 
order unity because the resolution measures contan the right power of 
$\pi$, see \cite{GCS}.
The bound depends on the point in phase 
space 
as expected. In order that the power of $t$ equals at least $k+1$
we should have $l\ge (12N-9)/(2(k+1))$. Thus we see that we need, for 
given $N$ the less terms in order to arrive at corrections of order 
$\hbar^{k+1}$ the higher is the power of 
$Q^2-<Q>^2$ in terms of which we expand the operator.\\ 
\\
Notice that the power counting argument outlined here can  
be supplemented by an analytical 
calculation in order to obtain the actual value of the integral because 
after the estimate of $<\psi_1,f\psi_2>$ everything is computable \cite{JBiP}. 
However, the leading power in $t$ of the integral will be at least the one 
we stated. Such an analytical calculation would also reveal all the 
finite, higher order (in $\hbar$) corrections that we have dropped in 
the power counting argument.

\section{Explicit Example: The case $N=2$}
\label{s3}

Notice that
for standard matter the terms appearing in the master constraint we have 
$2\le N\le 12$. We will exemplify the procedure for $N=2$ and consider 
arbitrary $k$. This means that in order to 
obtain a computable error bound of order $k+1$ we must have $l\ge 
15/(2(k+1))$. 

We will work out only the real part, the imaginary part works 
completely analogously. 
We have with $R=<\psi,p_1 f_1 p_2 f_2 p_3\psi>$ and 
$\psi_1:=\bar{p}_1,\;\psi_2:=p_2 f_2 p_3\psi$ 
\be \label{b.11a}
|\Re(R)-R_1|\le R_2 + R_3
\ee
where
\ba \label{b.12a}
R_1 &=& 
 \Re(<\psi, p_1 \bar{f}_1 p_2 f_2 p_3 \psi>)
\nonumber\\
R_2 &=& <\psi, p_1 [\Delta f]_1 \bar{p}_1 \psi>
\nonumber\\
R_3 &=& 
<\psi, \bar{p}_3 f_2 \bar{p}_2 [\Delta f_1] 
p_2 f_2 p_3 \psi>
\ea
The term $R_2$ is already computable by the methods of 
\cite{GCS}. The first term involves the non -- polynomial expression 
$f_2$ only linearly. Thus we can get rid of it by iterating the steps 
(\ref{b.11}) and (\ref{b.12}). Hence we set $\psi_1:=\bar{p}_2 
\bar{f}_1 \bar{p}_1 \psi,\;\;\psi_2:= p_3 \psi$ and find     
\be \label{b.13a}
|R_1 -R_4| \le R_5 +R_6 
\ee
where 
\ba \label{b.14a}
R_4 &= & 
\Re(<\psi, p_1 \bar{f}_1 p_2 \bar{f}_2 p_3 \psi>)
\nonumber\\
R_5 &=& 
<\psi, p_1 \bar{f}_1 p_2 [\Delta f_2] 
\bar{p}_2 \bar{f}_1 \bar{p}_1 \psi>
\nonumber\\
R_6 &:=& 
<\psi, \bar{p}_3 [\Delta f_2] p_3 \psi>  
\ea
All terms $R_4, R_5, R_6$ are computable by the methods of \cite{GCS}.

The term $R_3$ contains $f_2$ quadratically and iterating the steps 
(\ref{b.11}) and (\ref{b.12}) will not change that. However, we will 
iterate to generate terms of higher order. We set
$\psi_1=p_3 \psi,\;\psi_2:=\bar{p}_2 [\Delta f_1] p_2 f_2 p_3 \psi$ and 
get  
\be \label{b.15a}
|R_3-R_7| \le R_8 +R_9 
\ee
where
\ba \label{b.16c}
R_7 &= & 
\Re(<\psi, \bar{p}_3 \bar{f}_2 \bar{p}_2 
[\Delta f_1] p_2 f_2 p_3 \psi>)
\nonumber\\
R_8 &=& <\psi, \bar{p}_3 [\Delta f_2] p_3 \psi>
\nonumber\\
R_9 &:=& 
<\psi, \bar{p}_3 f_2 \bar{p}_2 [\Delta f_1] p_2 [\Delta f_2] 
\bar{p}_2 [\Delta f_1] p_2 f_2 p_3 \psi>  
\ea
The term $R_8$ is computable by the methods of \cite{GCS} while $R_7$ 
involves $f_2$ only linearly and which thus can be gotten rid of by 
performing 
once more steps (\ref{b.13}) and (\ref{b.14}). Hence we set 
$\psi_1=\bar{p}_2 [\Delta f_1] p_2 \bar{f}_2 p_3$ and
$\psi_2=p_3 \psi$ and find  
\be \label{b.16d}
|R_7-R_{10}| \le R_{11} +R_{12} 
\ee
where
\ba \label{b.16e}
R_{10} &= & 
\Re(<\psi, \bar{p}_3 \bar{f}_2 \bar{p}_2 
[\Delta f_1] p_2 \bar{f}_2 p_3 \psi>)
\nonumber\\
R_{11} &=& <\psi, \bar{p}_3 [\Delta f_2] p_3 \psi>
\nonumber\\
R_{12} &:=& 
<\psi, \bar{p}_3 \bar{f}_2 \bar{p}_2 [\Delta f]_1 p_2 [\Delta f_2] 
\bar{p}_2 [\Delta f_1] p_2 \bar{f}_2 p_3 \psi>  
\ea
All terms $R_{10},R_{11},R_{12}$ are computable by the methods of 
\cite{GCS}.

Notice that we have shown so far that
\be \label{b.16f}
|R-R_4|\le R_2+R_5+R_6+R_8+R_{10}+R_{11}+R_{12}+R_9
\ee
where only the last term $R_9$ is not computable. Since it contains 
three factors of the form $\Delta f$ meaning $l=1$ in order that it 
already leads to order $k+1$ corrections we must take $k=7$. 
For lower values of $k$ the procedure 
must be iterated again in order to generate more factors of 
the form $\Delta f$ in the non computable expressions. We will not do that 
as the 
general pattern should be clear by now. Notice that $R_2, R_5, R_6, R_8, 
R_{10}, R_{11}$ are of order $\hbar^{k+1}$ while $R_{12}$ is of order 
$\hbar^{3(k+1)}$ and thus is subleading.    

We finish this section by noticing that in  
the special case $N=2$ we actually do not need to use the iteration 
process in order to arrive at a calculable bound because we can use the 
Cauchy -- Schwarz inequality to obtain
\ba \label{b.16g}
&& |<\psi,p_1 f_1 p_2 f_2 p_3 \psi>| \le 
|| f_1 \bar{p}_1 \psi||\;\;
|| p_2 f_2 p_3 \psi||
\nonumber\\
& \le & 
|| f_1 \bar{p}_1 \psi||\;\;||f_2 p_3 \psi||
\nonumber\\
& \le & 
|| f^+_1 \bar{p}_1 \psi||\;\;||f^+_2 p_3 \psi||
\ea
where in the second step we used that $p_2$ is a bounded operator with 
bound $||p_2||\le 1$ and in the third we used $|| f 
\psi||^2=<\psi,f^2\psi>\le <\psi, (f^+)^2 \psi>\le ||f^+ \psi||^2$.
Both factors in the last line of (\ref{b.16g}) are computable by the 
methods of \cite{GCS}. This will suffice to get a bound of order $\hbar$
but unfortunately not better than that.

\section{Conclusion}
\label{s4}

In this paper we established that semiclassical calculations within AQG
and LQG can be done analytically as a perturbation expansion in $\hbar$.
The errors can be controlled, i.e. they are finite and can be estimated 
from above and are of higher order in $\hbar$ than the order that one is 
interested in. This is despite the fact that the spectrum of the volume 
operator of AQG is not known analytically, hence disproving the negative 
claims made 
in \cite{Nicolai} ``that nothing is calculable in LQG''. All one has to
do in order compute expectation values, correct to the order $\hbar^k$, 
with respect to coherent states
which involve the volume operator for a vertex $v$ in the form 
$(Q_v^2)^q$, is to replace this operator within the matrix element by 
its power expansion up to 
order $2k+1$ that is
\be \label{4.1}
<\psi,Q_v\psi>^{2q}[1+\sum_{n=1}^{2k+1}\; (-1)^{n+1}\;
\frac{q(1-q)..(2k-q)}{n!}\; (\frac{Q_v^2}{<\psi,Q_v\psi>^2}-1)^n]
\ee
The exact matrix elements of the operator $Q_v$ are known in closed 
form.
The error term can be estimated from above and is of order 
$\hbar^{k+1}$. Hence we are in a situation similar to ordinary QFT,
in fact the situation is even better because we have error control.

Formula (\ref{4.1}) is what one would naively do anyway in order to 
do semiclassical calculations, however, since the operator $Q_v$ is 
unbounded, one cannot use the spectral theorem to show that 
(\ref{4.1}) really approximates the actual operator. In this paper we 
showed that the naive guess is nevertheless correct by using methods 
from functional analysis and properties of our coherent states.\\
\\
Notice that while we have done perturbation theory within AQG for 
cubic algebraic graphs, everything goes through also in LQG for 
arbitrary graphs. Hence
the present paper turns AQG and LQG into a calculational discipline.\\ 
\\ 
\\ 
\\ 
{\large Acknowledgments}\\
\\
K.G. thanks the Heinrich B\"oll Stiftung for financial support. 
This research project was supported in part by a grant from NSERC of 
Canada to the Perimeter Institute for Theoretical Physics.

\end{document}